\documentclass[journal=jacsat,manuscript=article]{achemso}

\usepackage[version=3]{mhchem}
\usepackage{subcaption}
\usepackage{xcolor}
\usepackage{hyperref}

\newcommand{\revadd}[1]{#1}

\def\affIRCP{Chimie ParisTech, PSL University, CNRS, Institut de Recherche de Chimie Paris, 75005 Paris, France}
\def\affSorbonneU{Sorbonne Université, CNRS, Physico-chimie des Electrolytes et Nanosystèmes Interfaciaux, PHENIX, F-75005, Paris, France}
\def\notecontribution{These authors contributed equally to the work.}

\author{Emilio Méndez}
\affiliation{\affSorbonneU}
\altaffiliation{\notecontribution}
\author{Léna Triestram}
\affiliation{\affIRCP}
\altaffiliation{\notecontribution}
\author{Dune André}
\affiliation{\affIRCP}
\author{Fran\c{c}ois-Xavier Coudert}
\affiliation{\affIRCP}
\email{fx.coudert@chimieparistech.psl.eu}
\author{Rocio Semino}
\affiliation{\affSorbonneU}
\email{rocio.semino@sorbonne-universite.fr}

\title{\revadd{Systematic Study of Machine Learning Classification Algorithms of Zeolitic Imidazolate Framework Polymorphs}}

\begin{document}

\begin{tocentry}
\includegraphics[width=1\linewidth]{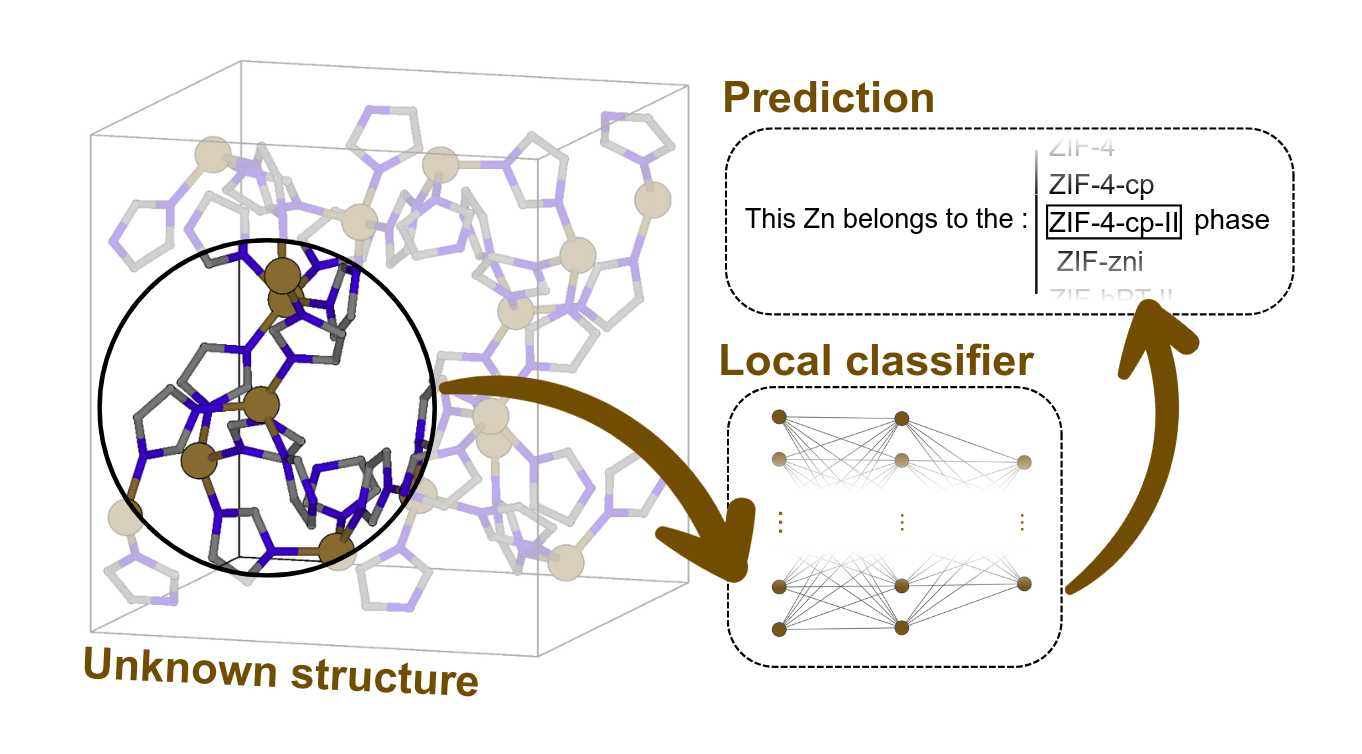}
\end{tocentry}

\begin{abstract}
Zeolitic Imidazolate Frameworks (ZIFs) are a family of metal--organic frameworks that feature metal centers tetrahedrally linked to imidazole-based ligands and adopt zeolite-like topologies. ZIFs formed by Zinc cations and imidazolate linkers exhibit a remarkable degree of polymorphism, which can be modulated by varying synthesis parameters or thermodynamic conditions (i.e., temperature and pressure). Computer simulations provide a unique way of studying these materials and their phase transitions from the microscopic standpoint, revealing their underlying molecular mechanisms. However, studying these mechanisms requires to be able to classify the phase of each molecular entity in an agnostic and automatic fashion, which is particularly challenging when the two phases involved are structurally very similar. In this work, we systematically study neural network classifiers to classify ZIF phases on-the-fly during molecular dynamics simulations. We test a variety of input features, differing both in the dimensionality and nature of the descriptors and in the kind of force field used for building the training/testing database. We reveal that even with low-dimensional descriptors the classification is highly accurate, 
while the use of high-dimensional descriptors leads to an even better performance.
Training the classifier with configurations coming from different force fields we can remove force field bias and enhance the classifier performance and general applicability. Finally, we apply our classifiers to reveal mechanistic details of the ZIF-4-cp $\xrightarrow{}$ ZIF-4-cp-II phase transition.
\end{abstract}

\section{Introduction}

Metal--organic frameworks, or MOFs, are hybrid organic--inorganic supramolecular assemblies based on relatively weak interactions that present a high number of intramolecular degrees of freedom. Due to this combination of factors, they present a higher propensity than classical solids for phenomena such as large-scale flexibility, framework dynamics, stimuli-responsive behavior, the presence of defects and disorder.\cite{Schneemann2014, Peh2020, Coudert2025} For the same reasons, many MOFs also display polymorphism, i.e., the ability to exhibit two or more crystalline phases that differ in their arrangement of the organic linkers or their conformation within the crystal lattice. Such MOFs can therefore display several phases with the same chemical composition but with differences in their structure, impacting or not the topology of the framework. As is known in the case of molecular crystals, polymorphism has an important impact on both the physical and chemical properties of MOFs,\cite{Aulakh2015} and has been widely studied both experimentally and computationally.\cite{Cheetham2018, CastilloBlas2024}

Zeolitic Imidazolate Frameworks (ZIFs) are a typical example for the occurrence of polymorphism in MOFs, with dozens of reported Zn(imidazole)$_2$ phases varying in topology as a function of synthesis conditions, linker functionalization, thermal and mechanical treatment, and other influences during and after their synthesis.\cite{Hughes2012, doi:10.1021/jacs.9b03234} This extends even to the existence of disordered phases, including liquids, glasses, gels, and composites thereof.\cite{Bumstead2020, Hou2020, Smirnova2023} However, despite the large amount of experimental evidence gathered on the polymorphism of ZIFs, many important questions remain widely open. The exact pressure--temperature phase diagram of ZIF-4, for example, has not yet been fully solved; although several studies have addressed the question, \cite{doi:10.1021/jacs.9b03234, D4TA05026F} these phase diagrams do not include many of the ZIF polymorphs originally found by Park and coworkers\cite{Park2006} nor are they really representative of thermodynamic stability, since it is well-known that the most stable phases are the dense ones, and porous phases occur due to kinetic considerations. Moreover, the nature and structure of some of the high-pressure phases of ZIF-4, ZIF-8 and ZIF-62 are still not fully solved,\cite{Vervoorts2021, Song2022, Robertson2025} the differences in the structures of the known phases can be quite subtle and some of the phases have a very narrow thermodynamic stability range.\cite{BousselduBourg2014} 

Computational techniques give us access to molecular-level resolution, which is key in revealing structural and mechanistic information pertaining to ZIF polymorphs and their phase transformations.\cite{D3DD00236E,castel_challenges_2022,mendez_microscopic_2024,D4TA05026F} However, simulation trajectories contain large volumes of data which require automation and agnostic data analysis methods to reveal all their hidden potential. For these reasons, there is interest in the development of computational methodologies for the identification of phases in ZIF systems. The challenge is even greater for the assignment of phases in a time-resolved and local manner. Being able to determine, during a molecular dynamics simulation, that a specific atomic environment at a given time is closest to a specific phase allows for the study of the dynamics of phase transitions and their mechanisms at the atomistic scale. This is made difficult because differences between structures are subtle, and simple geometrical criteria based on intuition are largely insufficient.\cite{Helfrecht2019} Techniques for local classification of phases have been employed previously: for example, geometric methods such as Polyhedral Template Matching (PTM)\cite{larsen_robust_2016} have been used to classify amorphous phases of silicon.\cite{rosset_signatures_2025}  In a previous study, a database of imidazolate-based ZIF polymorphs was created and used to train a neural-network classifier to identify ZIF polymorphs.\cite{mendez_microscopic_2024}

\revadd{In this work, we extend this previous methodology for local phase identification in ZIF systems in several directions. First, we broaden the classification problem from ambient-pressure phases to a larger family of crystalline and amorphous polymorphs spanning a wide range of thermodynamic conditions, including closely related high-pressure phases that are considerably more difficult to distinguish. Second, we investigate two complementary classes of local descriptors: Behler--Parrinello symmetry functions (BPSF)\cite{Behler2011} and Smooth Overlap of Atomic Positions (SOAP)\cite{Bartok2013}, differing both in dimensionality and in the structural information they encode, allowing us to assess the trade-off between computational efficiency and predictive accuracy. Finally, we investigate the effect of combining training data obtained from two different force fields, with the objective of improving the robustness and transferability of the resulting classifiers across interaction potentials.
Beyond the development of the classifiers themselves, we demonstrate their usefulness by applying them to the microscopic analysis of the ZIF-4-cp / ZIF-4-cp-II transformation. Unlike the crystal-to-amorphous and melting transitions investigated previously, this transformation occurs between two crystalline polymorphs with very subtle structural differences, providing a significantly more demanding benchmark for local phase identification. We identify a preferential direction in which the phase transition is slower, and correlate it to a change of relative orientation in the rings that are present in both closed-pore structures.}

\section{Systems \& Methods}

\subsection{ZIF-4 polymorphs and their classification}

We aim to develop robust methods of phase classification in ZIFs, which are able to identify the phases to which a Zn-centered environment belongs within a series of polymorphs containing both ordered and disordered phases. To do so, we present two phase-classifiers based on two general-purpose chemical descriptors.

\begin{figure}
    \centering
    \includegraphics[width=\linewidth]{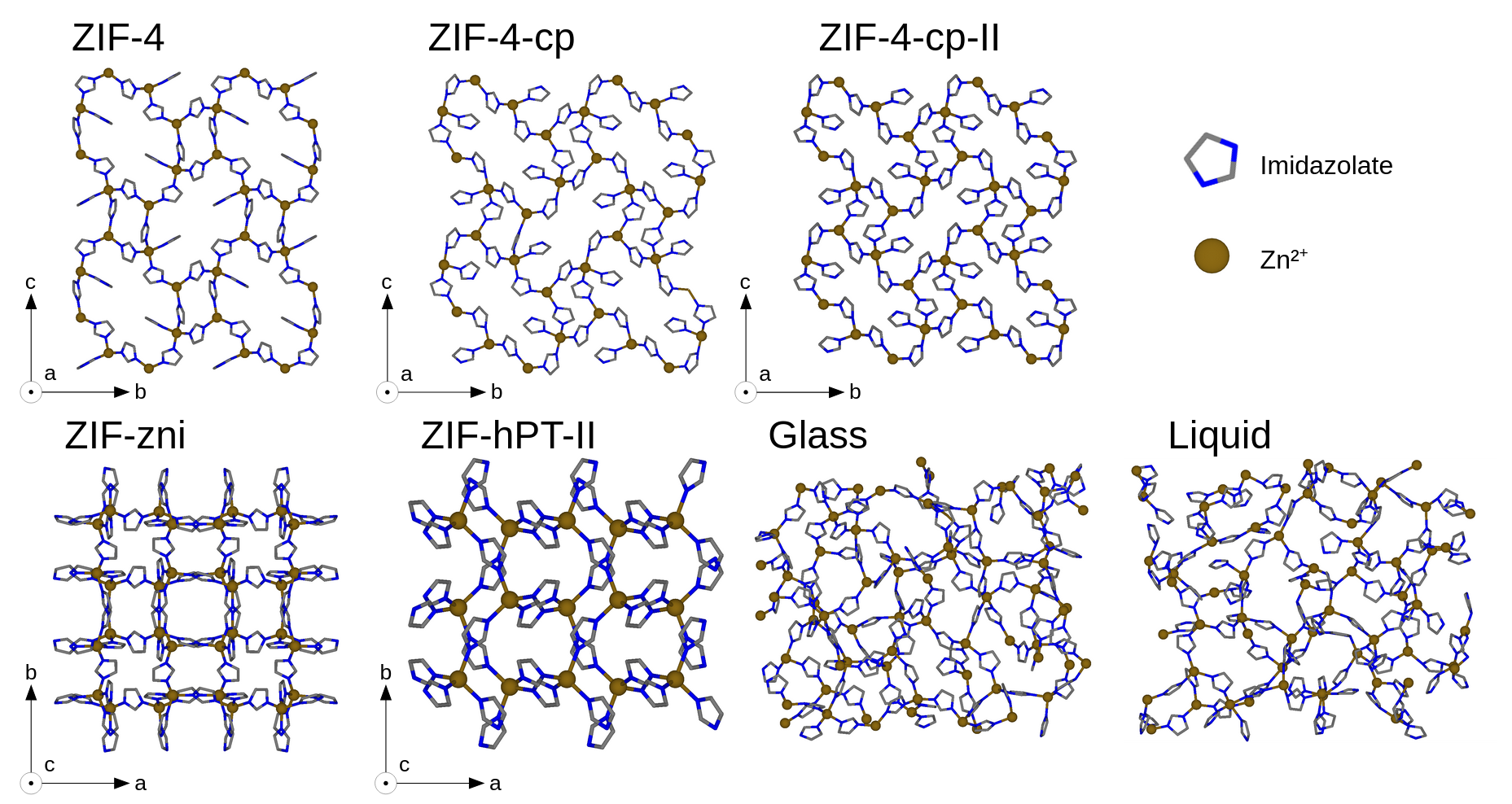}
    \caption{Structures of the polymorphs of ZIF-4 studied. Color code: Zn (ochre), N (blue), C (grey). Hydrogen atoms are not shown for clarity.}
    \label{fig:all_phases}
\end{figure}

We focused on seven distinct polymorphs with formula Zn(Im)$_2$, where Im stands for the unsubstituted imidazolate anion: C$_3$H$_3$N$_2^-$. These structures include five crystalline phases and two disordered phases which are shown in \autoref{fig:all_phases}. These polymorphs were previously described in both experimental\cite{doi:10.1021/jacs.9b03234,Henke2018} and theoretical\cite{D3DD00236E,castel_challenges_2022,mendez_microscopic_2024,D4TA05026F} works and can be obtained by varying the pressure and temperature (see the experimentally measured phase diagram in \autoref{fig:phasediagram} and in Ref. \citenum{doi:10.1021/jacs.9b03234}). Among the crystalline phases, three frameworks share the \emph{cag} topology: ZIF-4 is the open-pore phase, which undergoes a phase transition to the closed-pore phases when pressure is increased (ZIF-4-cp and ZIF-4-cp-II). The two remaining crystal phases ZIF-zni and ZIF-hPT-II have distinct topologies: \emph{zni} and \emph{dia} respectively. In all of these crystalline phases, the linkers adopt a tetrahedral geometry around the metal ion. We also include two phases that lack long-range order: a low-temperature glass phase and a high-temperature liquid phase.

\subsection{Simulation methods}

In order to build our classifiers, we first needed to create a database of configurations.
To do so, we performed MD simulations of each of the phases studied. To train generalizable classifiers, we produced configurations with two different kinds of force fields: the classical nb-ZIF-FF force field\cite{10.1063/5.0128656} and a machine-learning potential based on the MACE architecture.\cite{Triestram2026} 
On the one hand, the nb-ZIF-FF model is a partially reactive force field in which the Zn-Ligand bonds are modeled via Morse potentials, allowing the possibility of bond breaking and formation along a trajectory. The parameters of this model were optimized to reproduce experimental thermodynamic properties of different ZIF phases as well as \textit{ab initio}  Zn--ligand binding energies and it has been successfully applied to modeling phase transitions\cite{mendez_microscopic_2024,D4TA05026F} and self-assembly\cite{AndarziGargari2025,Mendez2025} of ZIFs. 
On the other hand, the machine learning potential is trained by fitting the energies and forces obtained from electronic structure calculations without imposing any connectivity at all. 
The potential was based on the MACE code\cite{batatia2022mace, batatia_mace_2023} (version 0.3.9). It was demonstrated in previous work by Triestam et al.\cite{Triestram2026} that this MLP provides a chemically accurate description of the different phases of ZIF-4, and is transferable across wide temperature and pressure ranges and has been trained on data sampled from the liquid phase of ZIF-4 through \emph{ab initio} molecular dynamics (AIMD).\cite{D3DD00236E} 
Since the two force fields have been constructed by very different methods, we consider that a mixture of configurations obtained by both methods may lead to model-agnostic classifiers.

We aim to classify the phases at a local Zn-centered environment level. In this way, each of the obtained configurations will include a number of data points equal to the number of Zn atoms present in the simulation.
Then, from the total set of configurations, we obtained 672000 Zn environments, 96000 for each phase (7 phases are considered) simulated with both force fields. In \autoref{dataset} we summarize the thermodynamic conditions at which each simulation was performed. These conditions were selected to lie inside the respective stability region of each phase. 

For the systems run with the ML force field, we performed the simulations under the NVT ensemble, because the force field does not reproduce the correct density in the NPT ensemble.\cite{Triestram2026}
The simulation details of each setup are summarized in the SI. \revadd{Using both NVT and NPT for the simulations allows us to obtain more diversity in the training data.}
\revadd{Since ZIF-4-cp and ZIF-4-cp-II exhibit very similar local environments, configurations corresponding to two thermodynamic conditions within the stability region of both phases were included in the training dataset. This broader sampling increases the structural variability during training and improves the accuracy of the discrimination between these closely related phases. The effect of this data augmentation in the performance of the classifier will be studied in the ``Case study: ZIF-4-cp/ZIF-4-cp-II transitions'' subsection of our Discussion.}

\begin{table}[ht]
\small
  \caption{Simulation conditions of temperature $T$, pressure $P$, density $\rho$ and number of Zn atoms $N_{\text{Zn}}$ in each configuration for the generation of the data set. For simulations in the NPT ensemble, the density corresponds to the average value along the trajectory.}
  \label{dataset}
  \begin{tabular*}{0.95\textwidth}{@{\extracolsep{\fill}}l|lllllll}
    \hline
    Phase & Forcefield  & Ensemble & $T$ (K) & $P$ (MPa) & $\rho$ (Zn/nm$^3$) & $N_{\text{Zn}}$ \\
    \hline
Liquid & nb-ZIF-FF & NPT & 700 & 0 & 4.05 & 64  \\
Liquid & MACE & NVT & 1500 & 803 & 3.65 & 128  \\
Glass & nb-ZIF-FF & NPT & 300 & 0 & 4.75 & 64  \\
Glass & MACE & NVT & 300 & 482 & 3.65 & 128  \\
ZIF-4 & nb-ZIF-FF & NPT & 300 & 0 & 4.03 & 128  \\
ZIF-4 & MACE & NVT & 300 & 633 & 3.65 & 128  \\
ZIF-4-cp & nb-ZIF-FF & NPT & 300 & 20 \& 105 & 4.76 \& 4.87 & 128  \\
ZIF-4-cp & MACE & NVT & 300 & 1218 \& 1494 & 4.70 \& 4.87 & 128  \\
ZIF-4-cp-II & nb-ZIF-FF & NPT & 300 & 150 & 5.11 & 128  \\
ZIF-4-cp-II & MACE & NVT & 300 & 1448 & 4.92 & 128  \\
ZIF-hPT-II & nb-ZIF-FF & NPT & 563 & 810 & 7.05 & 64  \\
ZIF-hPT-II & MACE & NVT & 560 & 7263 & 6.54 & 64 \\
ZIF-zni & nb-ZIF-FF & NPT & 300 & 0 & 4.92 & 128  \\
ZIF-zni & MACE & NVT & 300 & 508 & 4.43 & 64  \\
    \hline
  \end{tabular*}
\end{table}

\subsection{Classifiers}

To generate methods that allow to analyze structural transformations at a local level, we trained neural network classifiers based on local descriptors. As a result, we obtain as many predictions for a given structure as there are Zn atoms in the simulation box. This allows us to use the classifier in structures with different sizes and analyze processes at the atomic scale. These metal-centered predictions can also be averaged to yield global predictions, at the framework structure level.  

This approach was implemented in previous work by Méndez and Semino to classify ZIF-4, ZIF-zni, glass ZIF and liquid ZIF phases and the ZIF-4 $\longleftrightarrow$ glass phase transition.\cite{mendez_microscopic_2024} We now expand that methodology to include the rest of the ZIF phases from \autoref{fig:all_phases}, thus including structures that are stable at pressures higher than ambient pressure. We also aim to test the use of a more complete set of descriptors of Zn-centered environments and compare this method to a simpler version that is just the extension of that developed in Ref.~\citenum{mendez_microscopic_2024}, which relies on only 12 features per Zn atom. 

We thus generated two classifiers that differ in the nature and dimensionality of the descriptors used to encode the Zn environments.
One of them is based on BPSF, which are descriptors originally used for the development of machine learning potentials.\cite{Behler2011} This method has a low-associated computational cost since only twelve BPSF where selected to represent Zn-centered environments and only Zn neighbouring atoms within the environment were included in the spatial correlations, neglecting any ligand information. As a result, a vector of size 12 that encodes the spatial distribution of Zn atoms in the vicinity of the tagged Zn is obtained for each Zn atom in each configuration. The other classifier is based on the SOAP\cite{Bartok2013} descriptor. In contrast to the BPSF-classifier that only considers the metal ions surrounding the central Zn, the SOAP-classifier also considers the atoms which are part of the ligands and computes a 780-size vector, capturing more details of the environment.

The objective of this study is thus not to perform a direct comparison between SOAP and BPSF descriptors, but to present two different strategies for phase classification in ZIFs. Both BPSF and SOAP descriptors are rotationally and translationally invariant and their resulting vectors have a fixed size which does not depend on the number of neighbors in the local environment. 

\begin{figure}
    \centering
    \includegraphics[width=\linewidth]{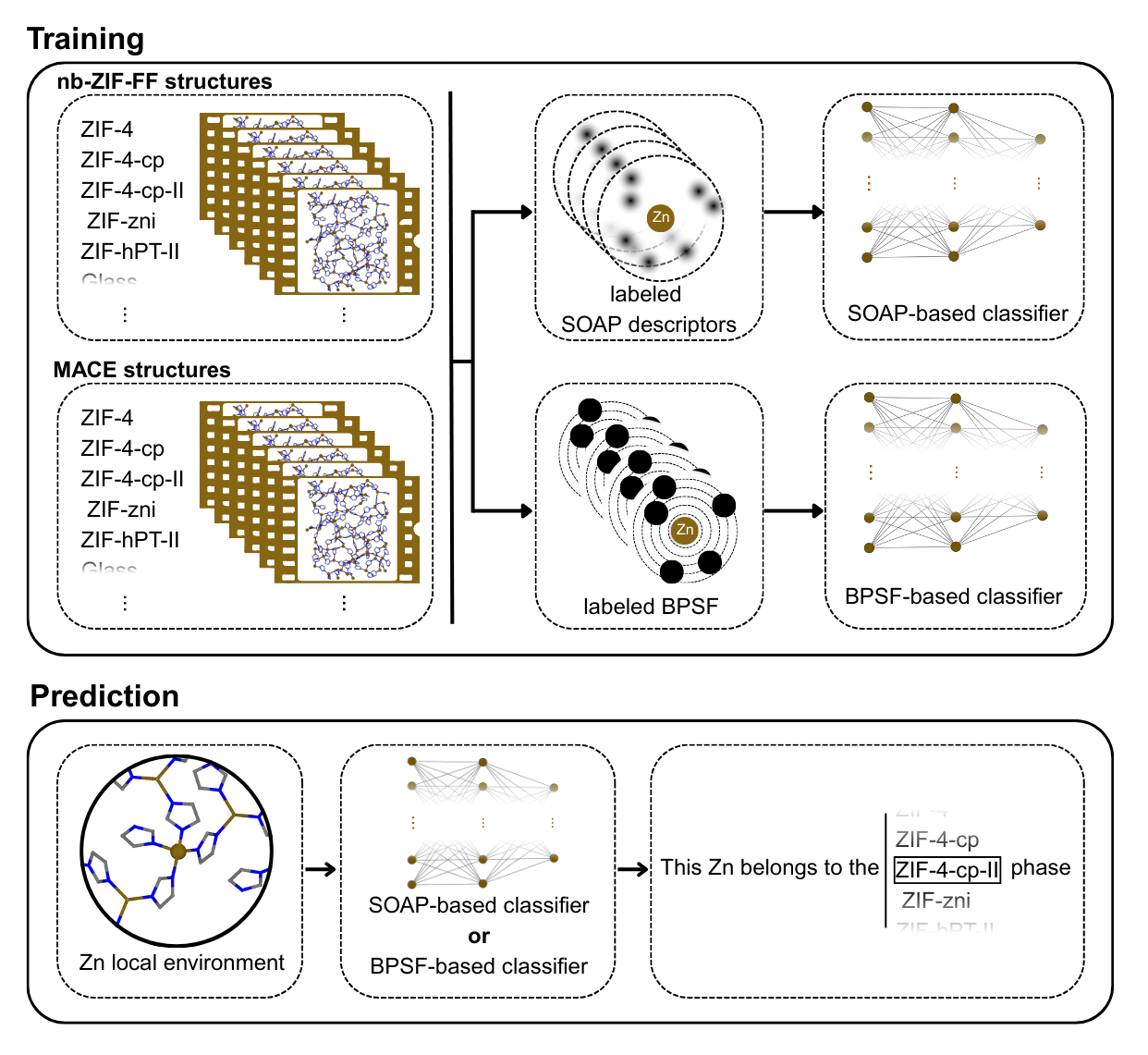}
    \caption{Workflow of the training and classification procedure for both classifiers. 
     Trajectories of ZIF-4 polymorphs were first generated from molecular dynamics simulations with two different force fields. SOAP and BPSF descriptors were then calculated to represent all Zn-centered environments. Subsequently, two neural networks were trained on these two sets of descriptors to generate two separate classifiers, which can then be used independently to classify the phase of a given Zn environment in a structure.}
    \label{fig:WORKFLOW}
\end{figure}

To train the classifiers, we used the database consisting of MD configurations of the seven phases presented in \autoref{dataset} and \autoref{fig:all_phases}. No distinction was made for the labels of the training data between the two simulation methods. To train the BPSF- and SOAP-based classifiers, we computed the respective descriptors for all the Zn-centers in the dataset  which results in feature vectors labeled according to their phase. For both classifiers, this data was then randomly split into training and test data in an {80-20\%} manner, and we trained two respective neural networks, which take the features as input and predict the phase label of the environment. The workflow for the training of the classifiers is represented in \autoref{fig:WORKFLOW}.

\revadd{Both classifiers employ fully connected feed-forward neural networks trained as multi-class classifiers. 
For the BPSF-based classifier, the neural network architecture comprised one input layer of 12 nodes, one for each symmetry function, a single hidden layer of 24 nodes with ReLU activation functions,\cite{agarap_deep_2019} and 7 output nodes with softmax activation functions.
For the SOAP-based classifier, the number of nodes in the first and second layers was set to 780 and 100 respectively.
The models were optimized using the Adam optimizer\cite{kingma_adam_2017} with categorical cross-entropy loss, while the training data were randomly divided into 80\% training and 20\% test sets. More details of each of the considered descriptors and the neural network training procedures are included in the SI.}

\revadd{To obtain a dimensionality reduction for the visualisation of the SOAP and BPSF descriptors, we use a similar methodology as described in Ref. \citenum{jorgensen_interpretable_2026}: we first do a pre-processing of the descriptor with PCovC\cite{jorgensen_interpretable_2026} to reduce the array to 5 dimensions, using a mixing parameter $\alpha$ of 0.5 and logistic regression. Subsequently, UMAP\cite{mcinnes_umap_2018} is applied to obtain a 2D representation. To make sure that our qualitative analysis remains unchanged with respect to the mixing parameter, we tested the impact of varying this parameter in \autoref{fig:SOAP_PCOVS5d_UMAP_alpha} and \autoref{fig:BPSF_PCOVS5d_UMAP_alpha}.}

\section{Results and discussion}
\subsection{Performance of the classifiers}

\begin{figure}
    \centering
    \includegraphics[width=\linewidth]{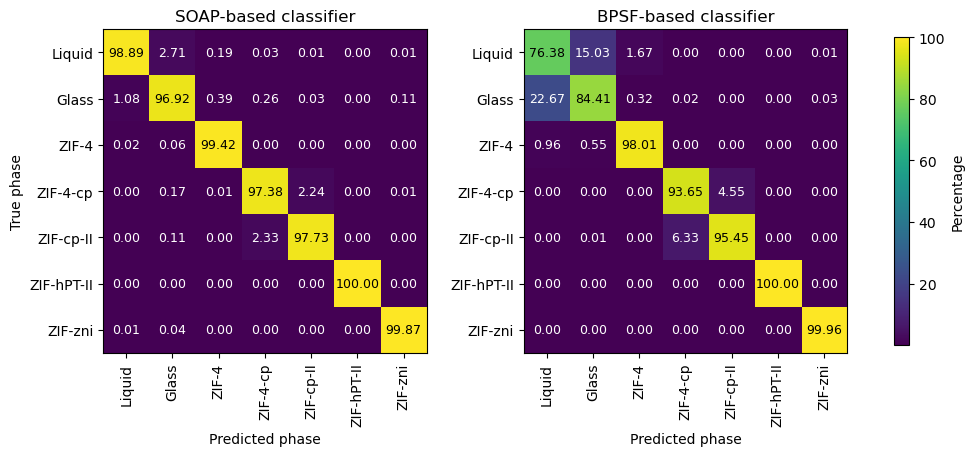}
    \caption{Confusion matrices of the SOAP- (left) and the BPSF-based (right) classifiers on the test set, which includes structures from both MACE and nb-ZIF-FF potentials. }
    \label{fig:confusion_matrix_full}
\end{figure}

Once trained on the local environments of the different phases, the SOAP- and BPSF-based classifiers achieve accuracies on the test set of 98.6\% and 92.6\% respectively. The confusion matrices of both classifiers are shown in \autoref{fig:confusion_matrix_full}. 

In a previous work\cite{mendez_microscopic_2024}, a different BPSF-based classifier which was trained on fewer phases (ZIF-4, ZIF-zni, glass and liquid) was presented, originating from nb-ZIF-FF simulations, which a classification accuracy of 90.3\% over the test set. The accuracy of the previous BPSF-based classifier is slightly lower because most of the confusion arises from the liquid/glass phases. By adding more data of these phases to the training dataset, the accuracy increases.

To analyze the confusion matrices and understand why some phases are better predicted than others, it is important to examine the structural differences between phases as well as the representation of the local chemical environments of the metal ion in these phases afforded by the two descriptors employed. The disordered phases (liquid and glass) and the two closed-pore phases (ZIF-4-cp and ZIF-4-cp-II) represent the largest source of error for both classifiers. This is expected from the structural similarities between these two pairs of phases. Both amorphous phases were already a source of error in the previous BPSF-classifier.\cite{mendez_microscopic_2024} Higher errors between the ZIF-4-cp and ZIF-4-cp-II phases were also expected: they have the same topology and very close atomic positions, which makes them challenging to differentiate.\cite{D4TA05026F} The ZIF-hPT-II structure, on the other hand, exhibits the best classification performance due to its distinct structure: its density is significantly higher than that of the other phases. 

We subsequently performed cross-experiments to assess force field dependence. For this, we trained new classifiers with the same methods on one set of structures (nb-ZIF-FF-generated structures) and tested them on a different set of structures (MACE-generated structures) and vice versa. The accuracy of the classifiers trained and tested with different force field data bases is detailed in \autoref{tab:accuracy}.
The resulting confusion matrices are represented in Figures S1 to S4 of the SI. The objective of these plots is to observe whether the classifiers were able to capture the physical features unique to each phase or if the classification is based on effects arising from the simulation methods, in other words, the objective is to analyze the generalizability of the classifiers.

\begin{table}[H]
    \centering
    \begin{tabular}{c|c|c|c}
           \hline
    Train set &\multicolumn{3}{c}{Test set} \\
         & nb-ZIF-FF & MACE &  All\\
       \hline
      nb-ZIF-FF & 99.5 (98.4)&76.4 (64.4)&87.9 (81.4) \\
      MACE   &  84.5 (64.9) & 98.6 (90.0) & 91.6 (77.4)\\
      All & 99.0 (98.0) & 98.2 (87.9) & 98.6 (92.6)\\
             \hline
    \end{tabular}
    \caption{Performance of the classifiers trained and tested with data sets produced by different force fields in \%. Results correspond to the SOAP-based classifier followed by BPSF-based classifier in parenthesis.}
    \label{tab:accuracy}
\end{table}

\revadd{The combination of both datasets for the training procedure significantly improves the performance of the classifiers.
The improved transferability obtained from mixed training datasets is consistent with the fact that each force field samples slightly different configurations of the same underlying structural motifs. Although both force fields reproduce the same crystalline phases, differences in the equilibrium densities, thermal fluctuations and simulation ensembles lead to variations in the local atomic environments. Training on both datasets therefore exposes the neural networks to a wider range of physically plausible configurations, reducing overfitting to the specific characteristics of a single force field and improving generalization when classifying previously unseen trajectories.}
For both cross-experiments (trained on MACE, tested on nb-ZIF-FF and trained on nb-ZIF-FF, tested on MACE), the amount of classification errors increases. This seems to be partly caused by the difference between the MACE and nb-ZIF-FF structures, and is correlated both to force field differences as well as to the use of different simulation ensembles.

\begin{figure}
    \centering
    \includegraphics[width=\linewidth]{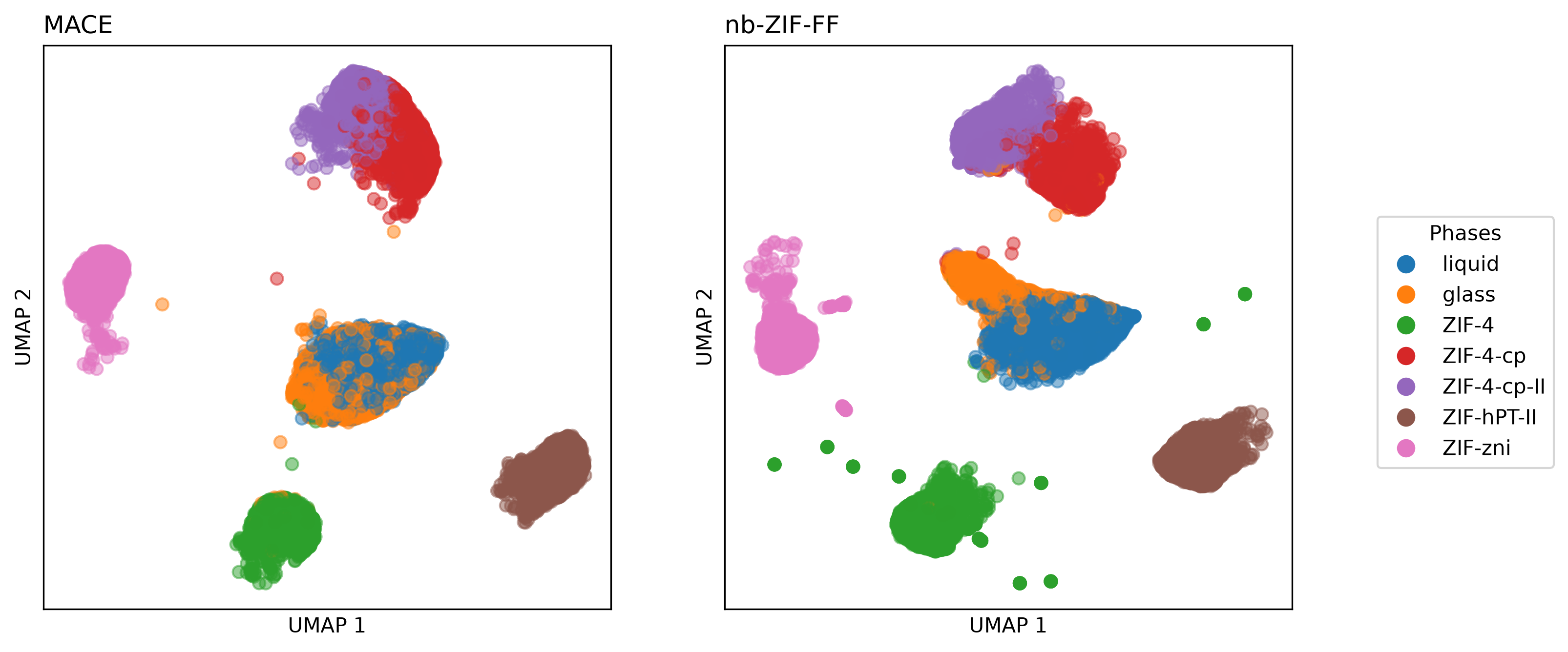}
    \caption{\revadd{UMAP projection with PCovC pre-processing of the Zn-centred SOAP descriptors for all studied phases, from molecular simulations with the MLP (left) and nb-ZIF-FF (right). The projection space is the same in both plots.}}
    \label{fig:SOAP_local_mace_nbZIFFF}
\end{figure}

In order to analyze visually the difference between phases as well as within the same phase modeled using the two force fields in the SOAP vector, we plot a 2-dimensional embedding of this high-dimensional vector in \autoref{fig:SOAP_local_mace_nbZIFFF}.
\revadd{The 2-dimensional representation is plotted in the same UMAP-space for both force fields (left and right panels). 
In this plot, the ZIF-4-cp and ZIF-4-cp-II descriptors cluster in the same region, which explains the confusion between these phases in the resulting classifier. This is also the case for the glass and liquid phases. For the SOAP descriptor, the glass phase seems to be the most different across force fields. 
We show a similar plot for the BPSF-based classifier in \autoref{fig:BPSF_PCOVSd_UMAP_2plots} which reveals that this classifier is more sensitive to density: for the MLP, the liquid, glass and ZIF-4 phases, which share the same density, are grouped in the same cluster. We can also notice larger disparities for the same phases across force fields. This might indicate that the BPSF-based classifier results in fewer chemically meaningful features than the SOAP-based one and thus the classification is more density-based. It is important to note that in these figures, we show only one possible 2D representation which captures the main features of these high-dimensional vectors.}

Although the BPSF-classifier does not contain ligand information and only uses a 12-feature descriptor, it is able to classify phases with high accuracy. The ligand information thus is not strictly necessary for the classification. However using a higher feature descriptor which takes into account the ligands seems to result in a more generalizable classifier and improve the accuracy.

The significant improvement in performance observed for classifiers trained on mixed force-field datasets, compared to those trained on individual force fields, suggests that combining data from multiple force fields is an effective strategy for improving generalization across simulation conditions.

\revadd{Compared with previous work\cite{mendez_microscopic_2024}, the present study systematically evaluates the robustness of local phase classifiers with respect to descriptor choice, simulation methodology and the complexity of the structural transformation under investigation. In particular, the distinction between the ZIF-4-cp and ZIF-4-cp-II polymorphs represents a considerably more demanding classification task than previously studied crystal–glass or crystal–liquid transformations\cite{mendez_microscopic_2024} because both phases share the same topology and differ only through subtle local structural distortions. Successfully resolving this transition demonstrates that the proposed framework is applicable to challenging polymorphic transformations beyond the cases previously reported.}

\subsection{Case study: ZIF-4-cp/ZIF-4-cp-II transitions}

We probed the capability of our classification algorithms to identify the phase transition between  ZIF-4-cp and  ZIF-4-cp-II (which we label CP and CPII respectively). This phase transition was selected for its high reversibility and for the close similarity between the structures of both phases, which makes the classification task more difficult than in other kinds of transitions. Furthermore, this transition could not yet be observed experimentally, although it was studied in a previous computational work \cite{D4TA05026F}. Nevertheless, the dynamic aspects of the growth of a CPII phase in the bulk of a CP phase and vise versa remain unknown.

For the analysis of the dynamic features of the CP/CPII interconversion, we generated non-equilibrium trajectories in which the system starts at one phase and is simulated in thermodynamic conditions in which the opposite phase is the most stable, so phase transition occurs spontaneously within a short simulation trajectory. Three independent simulations of CP to CPII transitions were generated and another three of CPII to CP to account for the stochasticity of the interconversion process. 
Since the transitions involve a change in volume, it was necessary to perform the simulations in the NPT ensemble, so the force field used to generate these transitions was nb-ZIF-FF, due to its higher stability in NPT simulations and lower computational requirements. The simulated phases comprise 6x6x6 unit cell systems in order to prevent periodic boundary conditions from affecting the process.
The simulation setup used for generating these transitions is detailed in the SI.
    
\begin{figure}
\begin{center}
\includegraphics[width=0.8\textwidth]{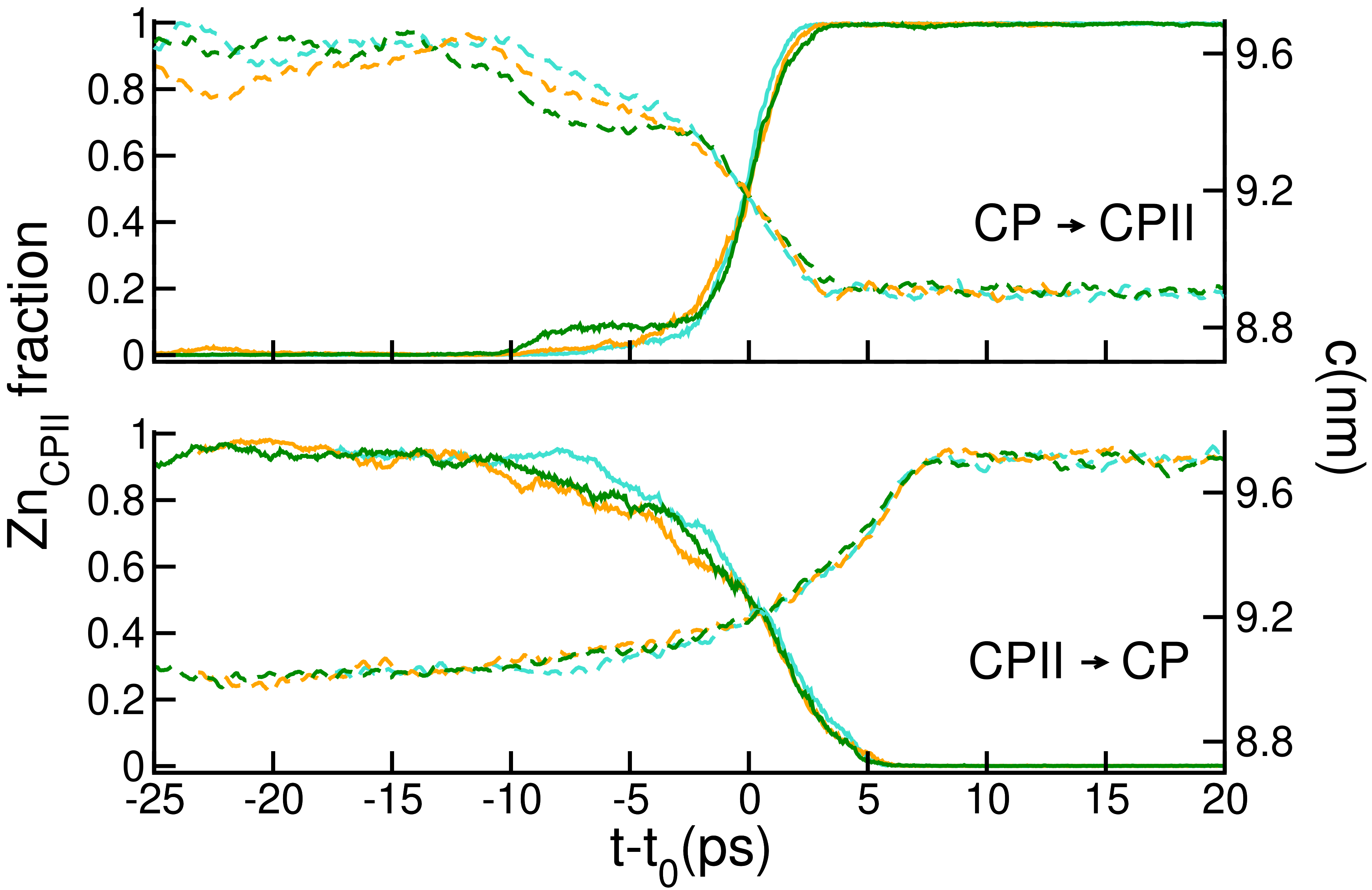}
\caption{Fraction of Zn ions classified as CPII by the SOAP-based algorithm (left axis, full lines) and simulation box size in the $z$ direction (along the crystallographic $c$ axis) (right axis, dashed lines) vs time for CP to CPII (upper panel) and CPII to CP (lower panel) transitions. All the trajectories are aligned at $t_0$, which corresponds to the time at which half of the Zn ions are classified as CPII. The different colors correspond to independent transition trajectories.} 
\label{fig:prediction}
\end{center}
\end{figure}

We applied the classification algorithm to each Zn ion in the system over time for each trajectory.
In \autoref{fig:prediction}, we plot the fraction of Zn atoms classified as CPII as a function of time, using the SOAP-based classifier, since it has a higher accuracy in the classification of these two phases. Equivalent plots for the BPSF-based classifier are included in the SI. In all cases, the fraction of Zn ions that were not classified as neither of the two phases was negligible.
Since the phase transition is an activated process, the absolute time at which it occurs is different in each trajectory. To align the plots from different simulations we centered each simulation at time $t_0$ in which the fraction of Zn atoms belonging to the final phase is 0.5. The plots that correspond to the BPSF-based classifier are centered using the same values of $t_0$ to allow a clear comparison.
We also plot, in dashed lines, the behavior of the $c$ cell parameter, which corresponds to the simulation box size in the $z$ direction.
It can be observed that the result of the classification algorithm is well correlated with the change in the $c$ cell parameter, which is characteristic of the  CP $\longleftrightarrow$ CPII phase transition. This demonstrates that the classifiers can detect the phase transition correctly.
In some trajectories, frustrated attempts of new phase formation are detected before the full phase transition occurs (see, for example, the behavior of the orange full curve in the upper panel at times between -25 and -20 ps). This kind of events can be associated to the nucleation of a new phase cluster that is not big enough to reach the critical size required to continue growing, and thus it reverts to the original phase.

The classification results present larger fluctuations before than after the phase transition occurs. This can be explained by the fact that the transition state of the system is metastable before the transition, and the training set of the algorithm was constructed only in thermodynamic conditions of stability. Also it could be attributed to the presence of small, fast-decaying clusters of the stable phase that nucleates in metastable conditions, as explained above.
Both classifiers are able to detect the phase transitions properly. The CP to CPII transitions are detected earlier in by the BPSF-based classifier while the opposite transitions are detected later. This means that the BPSF-based algorithm has a higher tendency to classify environments as CPII than the SOAP-one does. 
We also observe that the SOAP-based classifier presents higher fluctuations when applied to the metastable CPII phase, while the BPSF-based algorithm presents more fluctuations when applied to the metastable CP phase. This can be explained by the fact that the BPSF-based algorithm does not contain Zn--ligand spatial correlations, which makes it more sensitive to density fluctuations, which are higher in the CP phase. Again, the BPSF-based classifier overestimates the amount of Zn atoms in CPII phase, in line with the previous conclusions.

\revadd{In \autoref{fig:soap_without_cp_added} we present the same results of  \autoref{fig:prediction} but for the classifiers trained without the data that comes from structures of CP at $\rho = 4.87$  Zn/nm$^3$. The comparison between these two plots allows us to probe the influence of the thermodynamic conditions-based data augmentation in the performance of the classifiers.
For both classifiers, the transitions happen at slightly different times, as a consequence of the higher tendency to classify structures as CPII of the BPSF-based classifier. This is because the pressure conditions used for the generation of CP data are far from the transition pressure (i.e. 20 MPa versus 140 MPa\cite{D4TA05026F}).
Interestingly, the amount of fluctuations in the result of the SOAP-based classifier increases before the phase transition events. This corresponds to the situation in which a metastable phase dominates the simulation box. On the contrary, for the BPSF-based classifier, we observe an increase of the fluctuation always in the region where the CP phase dominates. The amount of misclassified structures in this case can go up to 60\% in the case of CP to CPII transitions and 30\% in the opposite one. This means that the BPSF-based classifier is more sensitive to the lack of CP data at pressures closer to the one at which the transition occurs. This is a consequence of the high sensitivity of this classifier to local density fluctuations as previously discussed, while the SOAP-based classifier, which takes into account ligand orientation information, produces a more robust result.
}

\begin{figure}
    \centering
    \includegraphics[width=\linewidth]{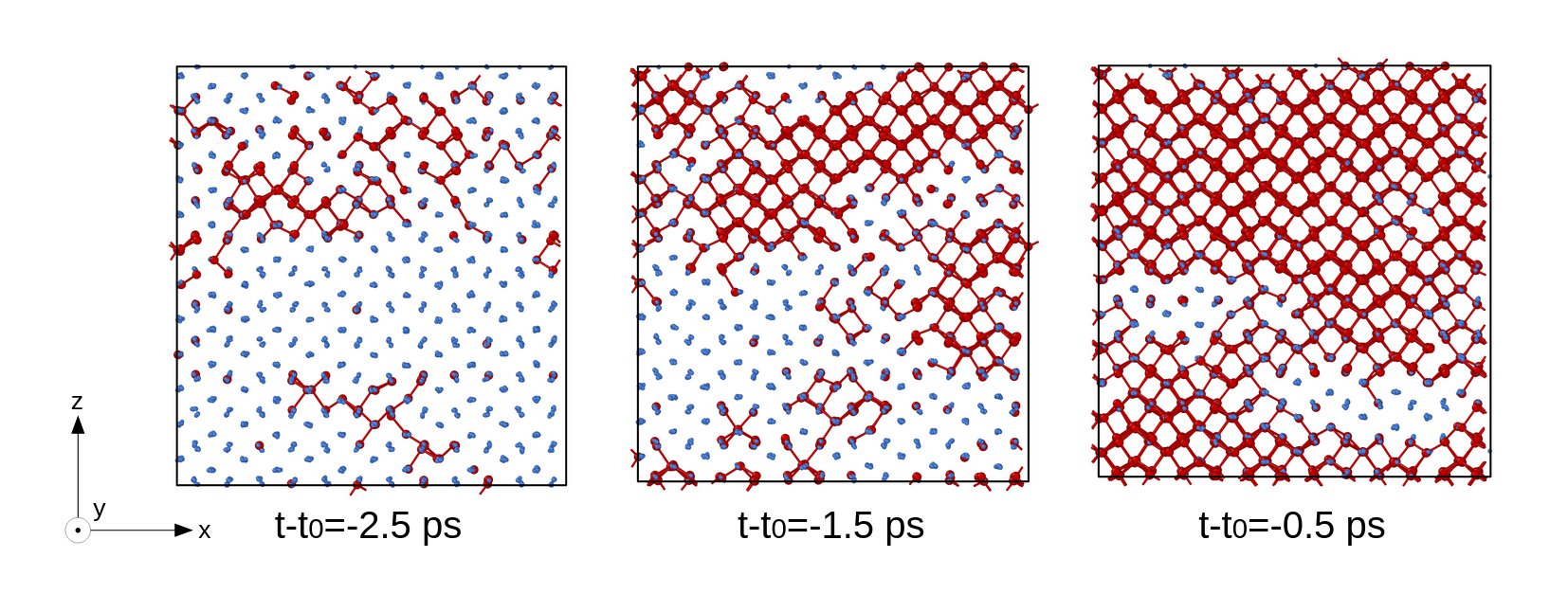}
    \caption{Snapshots of the phase transition from CP to CPII. Only Zn ions are represented and they are classified using the SOAP-based classifier, color-coded depending on the predicted phase, with CP environment in blue and CPII environment in red.}
    \label{fig:CP_to_CPII_1st_simulation}
\end{figure}

To provide a visual representation of the dynamics of the phase transition we plot in \autoref{fig:CP_to_CPII_1st_simulation} three snapshots of a trajectory in which a CP to CPII transition takes place. A similar plot for the \revadd{opposite} transition is included in the \revadd{\autoref{fig:CPII_to_CP_1st_simulation}}. The Zn atoms classified as CP are colored in blue while the ones classified as CPII are colored in red. In this picture we can observe how the new phase emerges from the bulk of the previous phase and expands until it occupies the full box. The atoms classified as CPII appear to be clustered and not randomly distributed. Also, we can observe the tendency of the new phase atoms to form in a layered-like fashion, parallel to the $(xy)$ plane. We will discuss these aspects more quantitatively below.

\begin{figure}
\begin{center}
\includegraphics[width=0.8\textwidth]{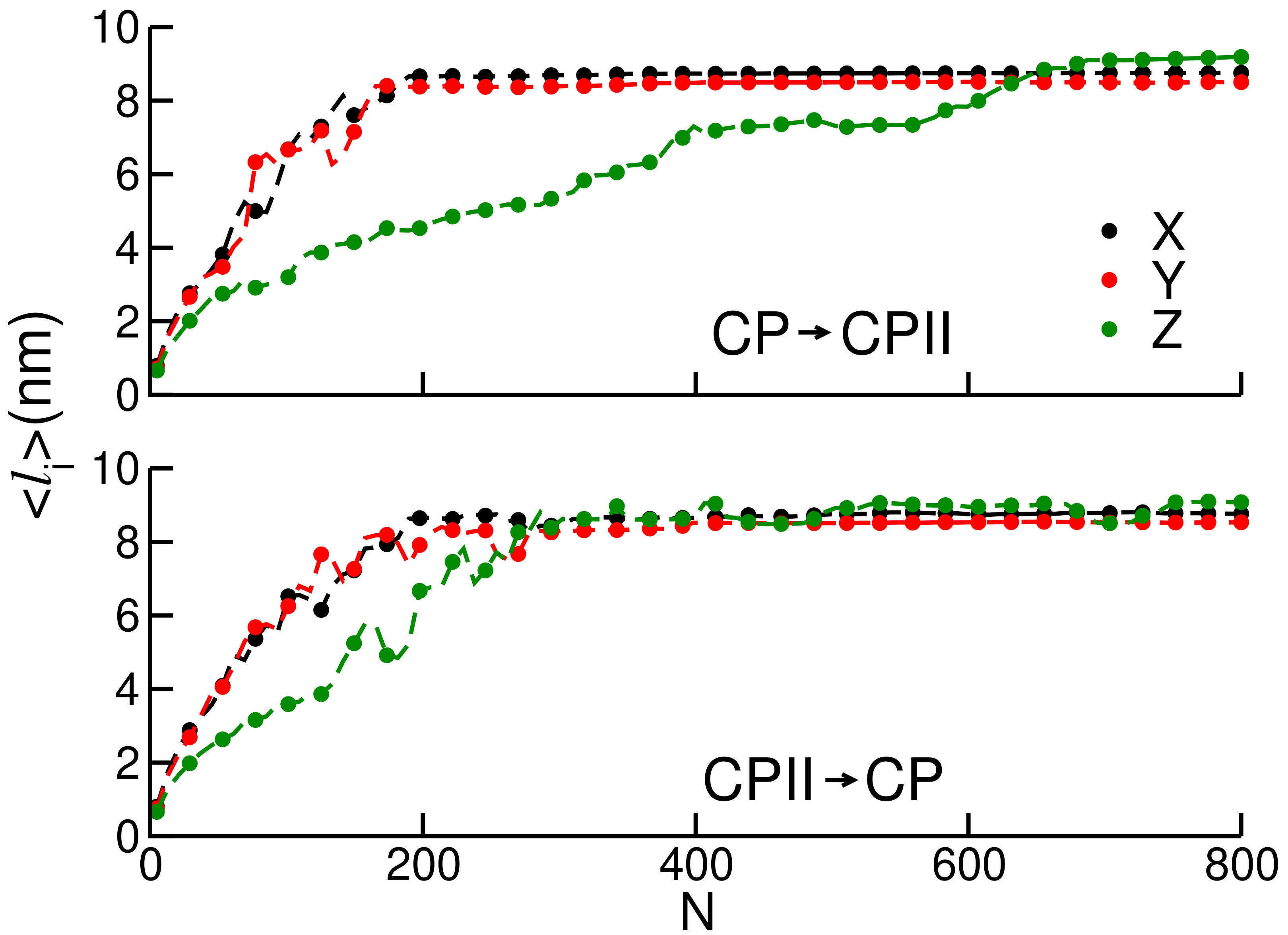}
\caption{Average cluster size $\langle l_i \rangle$ for directions $i = x$ (black), $y$ (red) and $z$ (green), plotted against the number $N$ of Zn ions comprising the cluster. Upper panel: CP to CPII transitions. Lower panel: CPII to CP transitions. The atoms that form the clusters correspond to the ones classified as belonging to the final phase in each case. Results from three independent simulations were averaged.} 
\label{fig:sizes}
\end{center}
\end{figure}

The capability of the algorithm to identify phases only using local information opens the possibility of analyzing geometrical features of the formation of the new phase. 
We performed a deep-first-search algorithm that identifies clusters of Zn ions that belong to the same phase according to the classifier. In the CP to CPII transitions we analyzed clusters of CPII Zn environments and the opposite in the CPII to CP transitions. In this way, we always considered clusters of the final phase formed.
In \autoref{fig:sizes} we plot the average size of the clusters composed of neighbor Zn ions that belong to the target phase. The size was measured as the largest distance in the three directions $x$, $y$ and $z$ among all pairs of atoms comprising the cluster. 
We observe that the growth of the new phase is non isotropic in all the transitions studied. For a given $N$, the size of the clusters is always bigger in $x$ and $y$ directions than in $z$. This means that the propagation of the new phase boundary over the simulation box is slower in the $z$ direction than in the other two. This behavior is also identified by the BPSF-based classifier, whose plot is shown in the SI. The visual inspection of the trajectory depicted in \autoref{fig:CP_to_CPII_1st_simulation} also confirms this trend.
The slower growth rate of the final phase in the $z$ direction can be attributed to the fact that the biggest change in cell parameters that occurs during a CP to CPII transition or vice versa is in the value of $c$, that corresponds to the size in the $z$ direction. This means that during the propagation of a phase boundary in the $z$ direction a greater change in linear density is produced, in comparison with a growth along $x$ or $y$.
This phenomenon seems to be more pronounced in the CP to CPII transition than in the CPII to CP one. Nevertheless, the opposite behavior is observed in the results that come from the BPSF-based algorithm. We can attribute this difference to higher error of the BPSF-based classifier in recognizing CP and CPII phases.

\begin{figure}
\begin{center}
\includegraphics[width=0.8\textwidth]{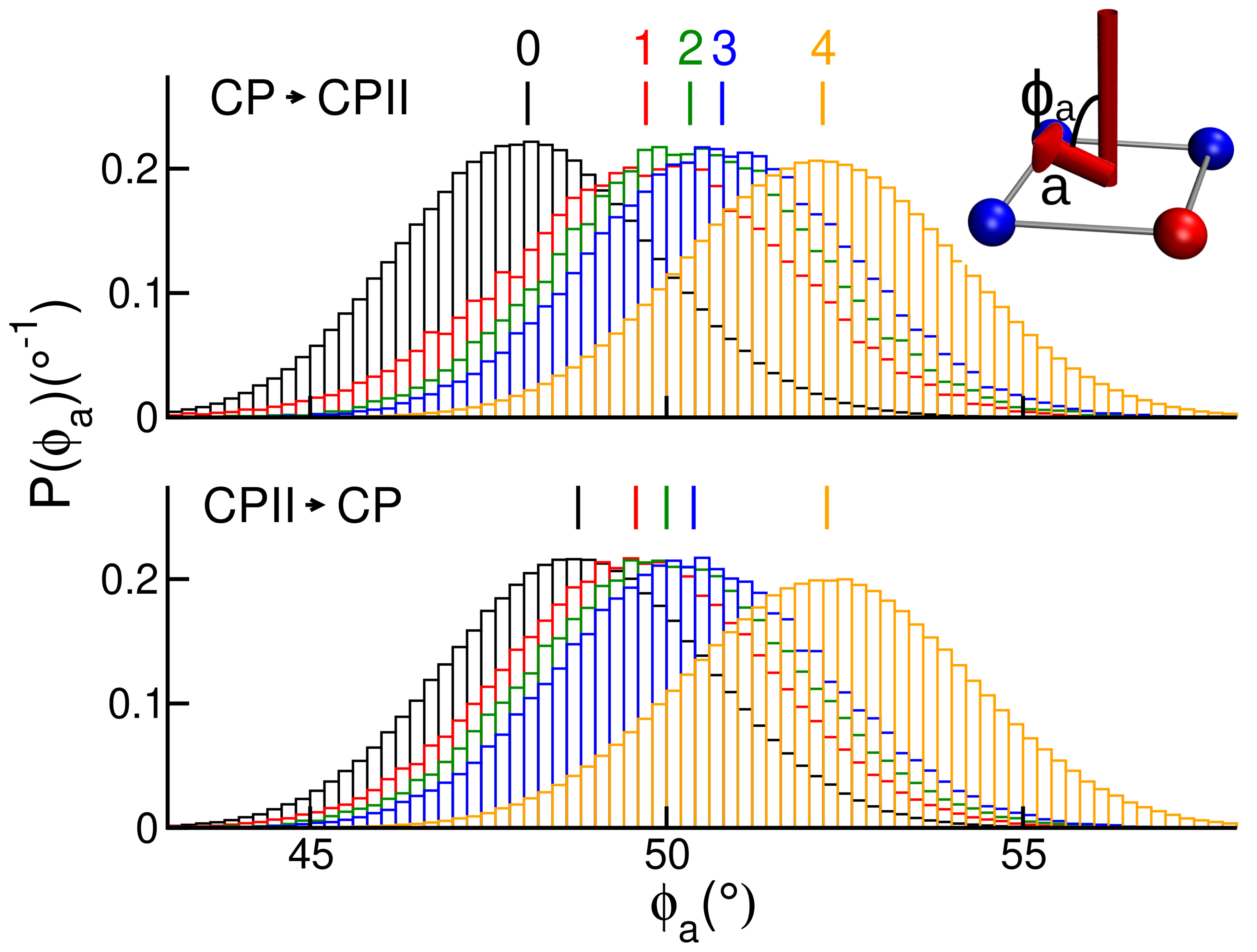}
\caption{Histograms of the angle $\phi_a$ between the vector normal to the surface formed by the four-membered rings of Zn atoms and the vector $\vec{a}$ that corresponds to the X axis of the simulation box. A sketch of the angle $\phi_a$ is depicted in the upper right corner, where each ball corresponds to a Zn atom forming a 4-membered ring, the vertical stick indicates the normal vector to the surface and the arrowed stick depicts the vector $\vec{a}$. Zn atoms are colored according to the classification result of the SOAP-based algorithm, blue for CP and red for CPII.
The histograms are divided by the number of Zn ions in the ring classified as CP. Mean values of each histogram are marked with vertical lines following the same color code.
Results obtained from transition trajectories of CP to CPII are plotted in the upper panel and the opposite ones in the lower panel.} 
\label{fig:tetragons}
\end{center}
\end{figure}

Finally, we sought to interpret the classifier output in physical terms by identifying which features of the local environments distinguish CP from CPII. 
To do so, we studied the correlation between the classification algorithm results and a local order parameter that that varies systematically across the transition.
Simple descriptors like average distances or angles between Zn and ligand species do not differ significantly between CP and CPII structures. For that reason we chose the orientation of the four-membered rings of Zn atoms, which had been identified in prior work as a local order parameter able to track the progress of the transition.\cite{D4TA05026F} To measure the orientation we computed the angle $\phi_a$ between the vector that is normal to the plane formed by the four-membered ring and the crystallographic vector $\vec{a}$ aligned with the X axis.
In \autoref{fig:tetragons} we plot the histograms of $\phi_a$ for each possible amount of Zn ions in the ring that are classified as CPII by the SOAP-based algorithm. A value of zero represents a ring fully classified as CPII, while a value of four represents a fully CP ring. In the upper right corner of the figure we show a sketch of the definition of $\phi_a$, which is explained in the caption. In all cases results from three independent trajectories were averaged.
For computing the normal vector of the plane, we calculated the normal vector to the planes formed by different triplets of atoms in the 4-membered ring and then averaged the result. In most cases, the four atoms almost lie in the same plane, so the values coming from individual triplets present small deviations with respect to the average. 
All the Zn ions in the crystal structure belong to at least one four-membered ring. In total there are four different ring orientations, but they all share the same histogram for the angle $\phi_a$, since they can be interconverted by reflection operations.
We can observe in the plot that there is a correlation between the average value of $\phi_a$ (marked with vertical lines) and the amount of Zn atoms classified as CPII, which means that the classifier is correctly capturing the change in local geometrical features that characterize the phase transition.\cite{D4TA05026F} The same trend is observed in the plots that correspond to the BPSF-based classifier, which are shown in the SI.
Nevertheless, $\phi_a$ cannot replace the classifier as order parameter to probe the phase transition, since there is a significant overlap in the histograms that correspond to each phase. 

The mean values of $\phi_a$ for each amount of CPII Zn ions are not the same in the the case where the transition goes from CP to CPII than in the reverse reaction. This is expected since the pressure values used in both simulations were different, changing the equilibrium values of all the geometrical parameters. Mild differences were detected between the two classification algorithms, being the mean values of $\phi_a$ closer to each other in the BPSF-based one, especially in the CP to CPII transition. This can be attributed to the fact that this classifier lacks the information about the bonding ligands orientation, which can improve the quality of the classification in cases where the structures of the two phases are similar.

\section{Conclusions \& Perspectives}

In this work, we developed and evaluated neural network classifiers for the identification of polymorph phases in ZIF-4 systems based on local atomic environments. By combining configurations generated from two distinct force fields and employing both low-dimensional (BPSF) and high-dimensional (SOAP) descriptors, we assessed the accuracy, robustness, and transferability of phase classification methods across a wide range of thermodynamic conditions.

Both classifiers achieve high predictive performance, with the SOAP-based approach yielding a higher accuracy. On the other hand, we show that the BPSF approach, even though containing only metal-centered information and a reduced amount of descriptors, can capture most of the relevant structural features required for phase discrimination. However, the inclusion of ligand information enhances robustness, particularly when comparing phases with similar structure, which are responsible for the greatest source of error.

\revadd{Training on configurations generated from two different force fields consistently improves predictive robustness relative to models trained on a single dataset. These results show that increasing the diversity of the sampled local environments enhances the transferability of the classifiers across different force fields.
}

Application of the classifiers to non-equilibrium molecular dynamics trajectories demonstrates their ability to resolve phase transitions in both space and time. The analysis of ZIF-4-cp $\longleftrightarrow$ ZIF-4-cp-II transitions reveals mechanistic aspects of the process. In particular, we detected an anisotropic growth behavior in $z$ direction with respect to the $x$ and $y$ axes.
We were also able to provide a physical interpretation of the classification result by correlating it with a microscopic order parameter that captures the transition between both phases. 
\revadd{The successful application of the algorithms to the challenging ZIF-4-cp $\longleftrightarrow$ ZIF-4-cp-II transition illustrates that local machine-learning classifiers can resolve subtle crystalline transformations and provide quantitative insight into nucleation and growth processes that are difficult to characterize using conventional structural descriptors alone.}

Future work could focus on the extension of the methodology to broader classes of ZIF-4 polymorphs that are missing in this analysis, such as ZIF-1, ZIF-3, ZIF-6 and ZIF-10\cite{Park2006} and their associated phase transitions, or to the classification of adsorption-induced structural transitions of flexible ZIFs.\cite{Tanaka2014, Chaplais2018} Overall, this study demonstrates that machine learning classifiers based on local descriptors provide a powerful and flexible framework for the automated analysis of polymorphism and phase behavior in metal–organic frameworks.

\begin{acknowledgement}
Access to high-performance computing platforms was provided by GENCI grants A0190807069 and A0170915688. This work was partly funded under the Imperial College London -- CNRS PhD joint programme. This work was partly funded by the European Union ERC Starting grant MAGNIFY, grant number 101042514.
\end{acknowledgement}

\begin{suppinfo}
Full simulation details, analysis of force field dependence of the classifiers performances, molecular snapshots, BPSF-based classifier results for ZIF-4-cp/ZIF-4-cp-II interconversions, experimental phase diagram of ZIF-4.
\end{suppinfo}

\section*{Data and Software Availability}
In order to make our work fully reproducible by others, representative input files for each type of simulation are available online in our data repositories at \url{https://github.com/rosemino/MAGNIFY/tree/main/ZIFs_NN_classifiers}.

\section*{Author Contributions}
F.-X. C. and R. S.: Conceptualization, Methodology, Investigation, Funding acquisition, Resources, Supervision, Writing – original draft, Writing – review \& editing. E. M., L. T. and D. A.: Conceptualization, Methodology, Investigation, Data curation, Formal analysis, Software, Visualization, Writing – original draft, Writing – review \& editing.

\section*{Conflict of Interest statement}
The authors declare no competing financial interests.


\newpage
\setcounter{figure}{0}
\setcounter{table}{0}

\renewcommand{\thesection}{S\arabic{section}}
\renewcommand{\thetable}{S\arabic{table}}
\renewcommand{\thefigure}{S\arabic{figure}}
\renewcommand{\theequation}{S\arabic{equation}}

\section{Supporting Information for:\\ Systematic Study of Machine Learning Classification Algorithms of Zeolitic Imidazolate Framework Polymorphs}

\section{Simulation details}

\subsection{nb-ZIF-FF force field simulations}

To generate configurations for the training set with the nb-ZIF-FF force field\cite{10.1063/5.0128656} we performed molecular dynamics simulations in the NPT ensemble. The MD timestep was set to 0.5~fs. Configurations were taken every 10,000 simulation steps to avoid correlations. Nosé--Hoover thermostats and barostats were employed, with a decaying time of 50~fs and 500~fs respectively. All simulations were performed using the LAMMPS molecular dynamics software.\cite{plimpton_lammps_2023}
ZIF-4, ZIF-zni and ZIF-hPT-II initial configurations were obtained from experimental CIF files\cite{doi:10.1021/jacs.9b03234}. Other phases were obtained starting from ZIF-4 by melt-quench protocols (for the liquid and glass) or by changes in pressure (for ZIF-4-cp and ZIF-4-cp-II) as done in previous works\cite{D4TA05026F,mendez_microscopic_2024}. Each structure was properly equilibrated before production runs.

\subsubsection{ZIF-4-cp - ZIF-4-cpII transition trajectories}
The thermodynamic parameter used to control the relative stability of ZIF-4-cp and ZIF-4-cp-II phases was the pressure. We know from previous studies that for the nb-ZIF-FF potential, the coexistence pressure of ZIF-4-cp and ZIF-4-cp-II is 145 MPa at ambient temperature.\cite{D4TA05026F}
For the generation of trajectories in which ZIF-4-cp to ZIF-4-cp-II transitions take place to probe the classifiers performance, we implemented the following procedure. We started from a ZIF-4-cp system equilibrated at 120 MPa, at which ZIF-4-cp is the most stable phase. Then, we performed several simulations in which the pressure was increased by 5 MPa smoothly during 10$^5$ steps and re-equilibrated for another 10$^5$ steps. With this procedure, we were able to obtain a metastable ZIF-4-cp state at 160 MPa, where the most stable phase is ZIF-4-cp-II. Subsequent increases in pressure lead to instantaneous transitions to ZIF-4-cp-II. 
Starting from this configuration, we run a long simulation at 160 MPa until a transition to ZIF-4-cp-II was observed and the new phase was stabilized. Transitions are identified by abrupt changes in the cell parameters of the system.
In order to have better statistics, the procedure was repeated three times with different initial configurations, so that we end up with three equivalent independent transition trajectories.
The same scheme was used to obtain trajectories in which a ZIF-4-cp-II to ZIF-4-cp transition occurs. In this case, we started with a ZIF-4-cp-II phase at 170 MPa and decreased smoothly the pressure until a value of 130 MPa was reached. This value was chosen in such a way that the difference in pressure with respect to the coexistence value is the same as in opposite transition ($145 \pm 15$~MPa). By doing so, we ensure that the driving force of the transition is comparable in both cases. 
The simulated systems comprised $6\times 6\times 6$ unit cells, which represents a total of 86400 atoms. 

\subsection{Machine learning force field simulations}

For the generation of the atomic structures, we performed MD simulations in the $(N, V, T)$ ensemble, with a timestep of 0.25~fs for the liquid phase and 0.5~fs for the remaining phases. The glass and liquid configurations were generated using the melt-quench protocol from ref \citenum{Triestram2026} with the ZIF-4 starting structure from Widmer et al.\cite{doi:10.1021/jacs.9b03234} The ZIF-hPT-II, ZIF-zni and ZIF-4-cp-II starting structures are the same as for the nb-ZIF-FF simulations. Due to the lack of experimental structure, the ZIF-4-cp structure was taken as the average structure of the ZIF-4-cp nb-ZIF-FF simulation by averaging the lattice parameters and atomic positions over the simulation. Simulations were run on a $2\times 2 \times 2$ supercell for all phases except for the ZIF-zni structure for which a $1\times 1 \times 2$ supercell was used. We employ a Nos\'e--Hoover thermostat and barostat with a damping factor of 0.1~ps for the thermostat and 1~ps for the barostat. Prior to production simulations and data gathering, the structures were equilibrated.

The simulations were performed using LAMMPS\cite{plimpton_lammps_2023} (version 29 Aug 2024) using the \texttt{LAMMPS\_MACE} model that relies on KOKKOS\cite{LAMMPS_KOKKOS, KOKKOS3} to offload calculations to the GPU --- in this case, a NVIDIA Tesla V100. 

\begin{table}[H]
    \centering
    \begin{tabular}{c|c|c|c}
      coordination number   & ZIF-4 & glass &  liquid\\
       \hline
      \emph{ab initio}   & 4 &3.93 & 3.52 \\
      MLP   &  4 & 3.92 & 3.54 \\
    \end{tabular}
    \caption{Zn-N coordination numbers for ZIF-4, the liquid and the glass phases for the MLP and \textit{ab initio} simulations.}
    \label{tab:placeholder}
\end{table}

\section{Descriptor details}

\subsection{SOAP-based Classifier}

The first classifier is based on the local Smooth Overlap of Atomic Positions (SOAP) vector which is a local descriptor of the atomic environment of a given atom. Here, we calculate it only centered on Zn ions, however, SOAP takes all the atoms of the local environment into account up to a cutoff radius, including the atoms of the linker.  
To calculate these vectors, we used the DScribe library, version 2.1.1. We computed the SOAP vectors of 96000 Zn-centers for each of the seven phases. 
SOAP vectors were calculated in a Gaussian basis with a cutoff of 9 \AA. We used a maximum angular degree for the spherical harmonics of 9 and 12 radial basis functions.

We used the Multi-layer Perceptron classifier from the scikit learn library (version 1.6.1)\cite{scikit-learn} with one hidden layer of 100 neurons using Rectified Linear Unit (ReLU) activation functions\cite{agarap_deep_2019}. We used the Adam optimizer\cite{kingma_adam_2017} and a maximum of 1000 epochs for training. The output is a single scalar designating the predicted phase. 

\subsection{BPSF-based Classifier}

For the Symmetry Functions-based classifier, we computed 12 BPSF for each Zn environment in the data set, 4 of them comprised radial functions, while the other 8 comprised angular ones. The explicit expressions of these functions are given in Eqs.\ref{eq:sfrad} and \ref{eq:sfang} respectively. In these equations, $R_{ij}$ is the distance between atoms $i$ and $j$, $\theta_{ijk}$ is the angle between atoms $j$, $i$ and $k$ in that order, and $f_c$ is a cutoff function that decays to zero at a distance $R_c$. $\eta$ and $\lambda$ are parameters that change between different symmetry functions. Only Zn atoms were considered for the environment calculations, both as centers and within the environment. 

\begin{eqnarray}
G_i^{rad}=\sum_{j \neq i} e^{-\eta R_{ij}^2} \cdot f_c(R_{ij}) 
\label{eq:sfrad}
\end{eqnarray}

\begin{eqnarray}
G_i^{ang}=\sum_{j,k\neq i} (1-\lambda \cos{\theta_{ijk}}) \cdot e^{-\eta(R_{ij}^2+R_{ik}^2+R_{jk}^2)} 
\cdot f_c(R_{ij}) \cdot f_c(R_{ik}) \cdot f_c(R_{jk})
\label{eq:sfang}
\end{eqnarray}

The values of $\eta$ for the radial functions were 0.005, 0.0075, 0.01 and
0.02 Bohr$^{-2}$. With each of these $\eta$ values, we construct two angular functions with $\lambda$ $=\pm$1. The
cutoff radius $R_c$ was set to 13 \AA, comprising approximately the first two coordination shells.
The symmetry functions were normalised so that the highest and lowest values in the database
were assigned one and zero respectively. 

The neural network architecture comprises one input layer of 12 nodes, one for each symmetry function, a single hidden layer of 24 nodes with ReLU activation functions, and 7 output nodes with softmax activation functions, which guaranty that the outputs are positive numbers that sum to one.
The categorical cross-entropy was used as loss function to be minimized during the training. We employed the Adam algorithm for the optimization
of the neural network parameters. This process was done over 350 epochs of the training set.

\begin{figure}
    \centering
    \includegraphics[width=\linewidth]{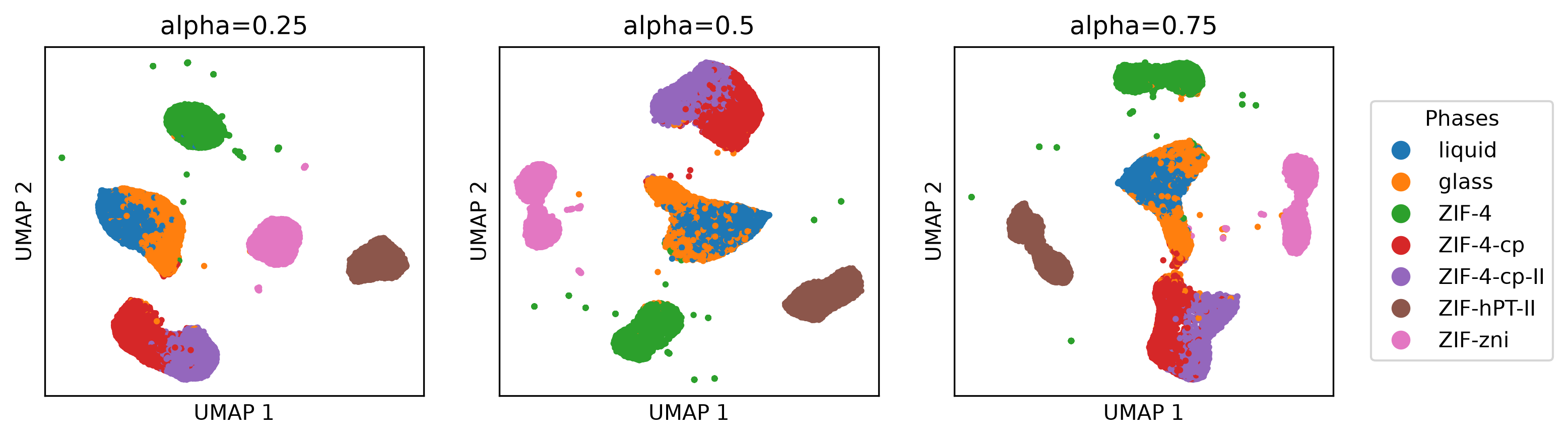}
    \caption{Mixing parameter-dependence of the UMAP projection with PCovC pre-processing of the SOAP descriptors for all studied phases.}
    \label{fig:SOAP_PCOVS5d_UMAP_alpha}
\end{figure}

\begin{figure}
    \centering
    \includegraphics[width=\linewidth]{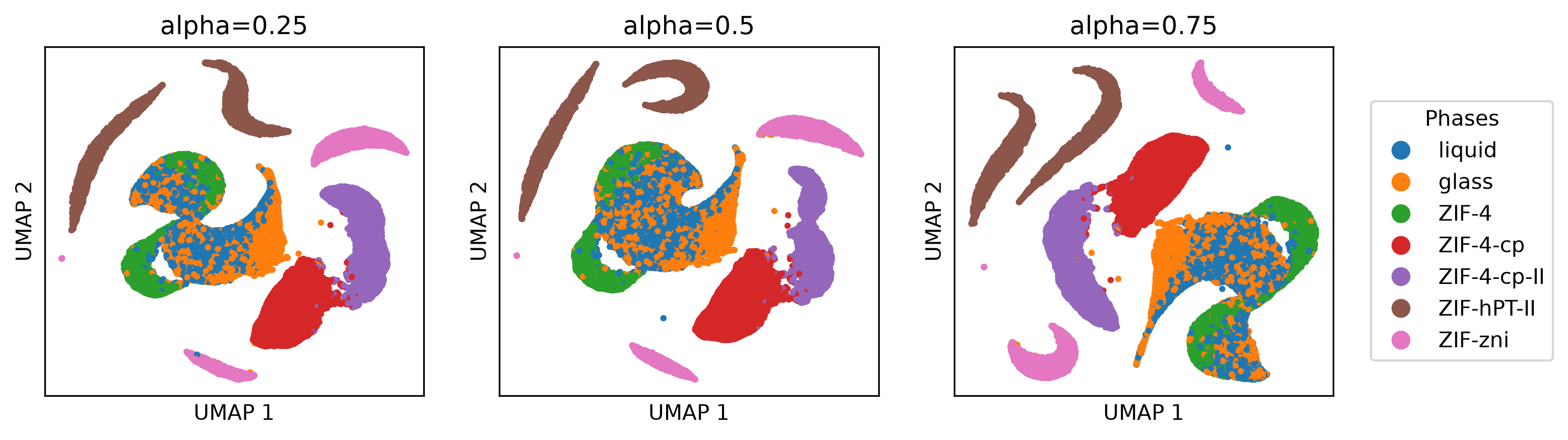}
    \caption{Mixing parameter-dependence of the UMAP projection with PCovC pre-processing of the BPSF descriptors for all studied phases.}
    \label{fig:BPSF_PCOVS5d_UMAP_alpha}
\end{figure}

\section{Force field dependence of the classifiers performances}

\begin{figure}
    \centering
    \includegraphics[width=\linewidth]{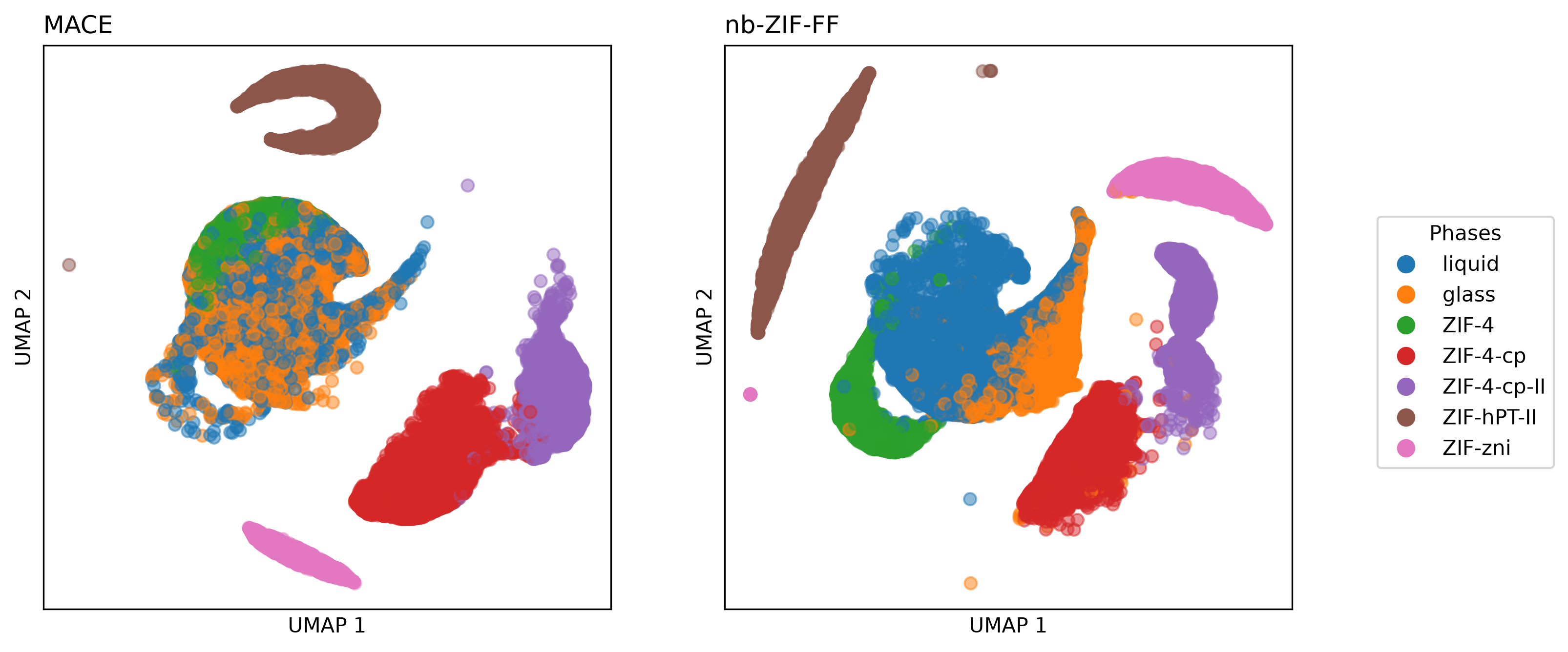}
    \caption{UMAP projection with PCovC pre-processing of the BPSF descriptors for all studied phases, from molecular simulations with the MLP (left) and nb-ZIF-FF (right).The projection space is the same in both plots.}
    \label{fig:BPSF_PCOVSd_UMAP_2plots}
\end{figure}

\begin{figure}[H]
    \centering
   \includegraphics[width=1\linewidth]{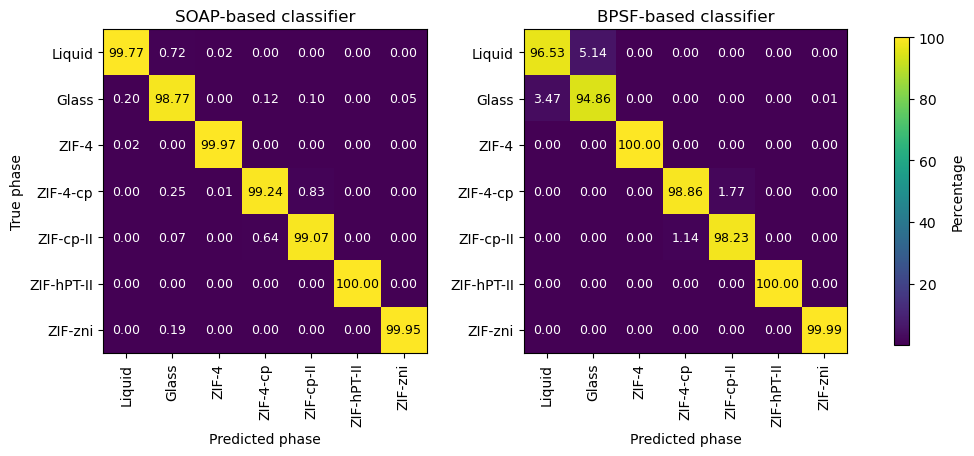}
    \caption{Confusion matrices of the SOAP- (left) and the BPSF-based (right) classifiers trained solely on structures from the nb-ZIF-FF potential and tested on structures from the test set of the same potential.}
    \label{fig:confusion_matrix_nb_on_nb}
\end{figure}

\begin{figure}[H]
    \centering
   \includegraphics[width=1\linewidth]{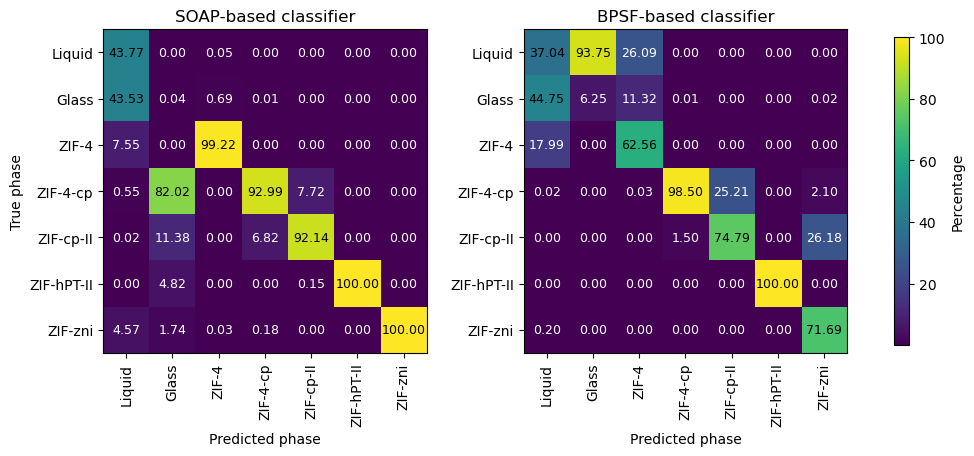}
    \caption{Confusion matrices of the SOAP- (left) and the BPSF-based (right) classifiers trained solely on structures from the nb-ZIF-FF potential and tested on structures from the MACE potential.}
    \label{fig:confusion_matrix_nb_on_nn}
\end{figure}

\begin{figure}[H]
    \centering
   \includegraphics[width=1\linewidth]{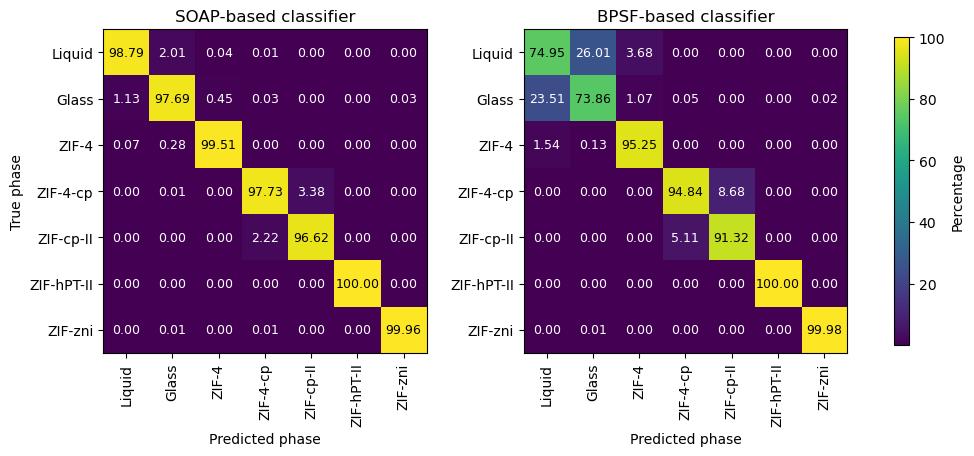}
    \caption{Confusion matrices of the SOAP- (left) and the BPSF-based (right) classifiers trained solely on structures from the MACE potential and tested on structures from the test set of the same potential.}
    \label{fig:confusion_matrix_nn_on_nn}
\end{figure}

\begin{figure}[H]
    \centering
   \includegraphics[width=1\linewidth]{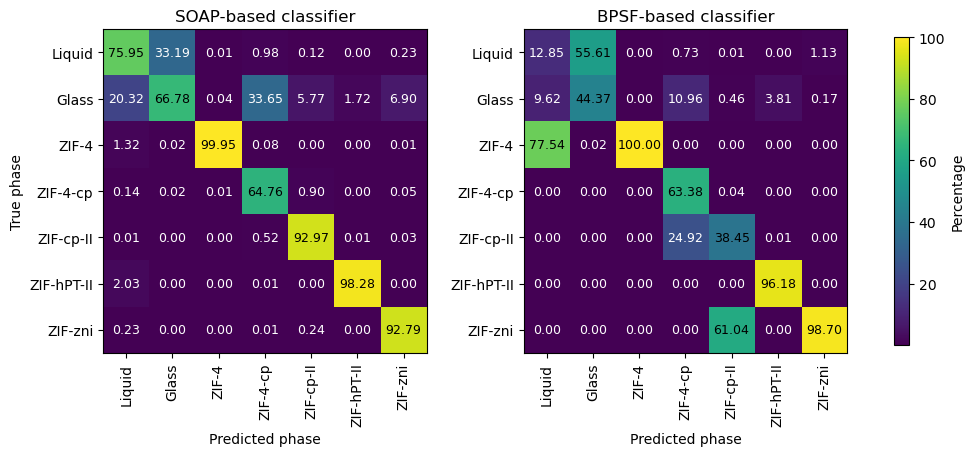}
    \caption{Confusion matrices of the SOAP- (left) and the BPSF-based (right) classifiers trained solely on structures from the MACE potential and tested on structures from the nb-ZIF-FF potential.}
    \label{fig:confusion_matrix_nn_on_nb}
\end{figure}

\section{ZIF-4-cp-II to ZIF-4-cp snapshots}

\begin{figure}[H]
    \centering
    \includegraphics[width=\linewidth]{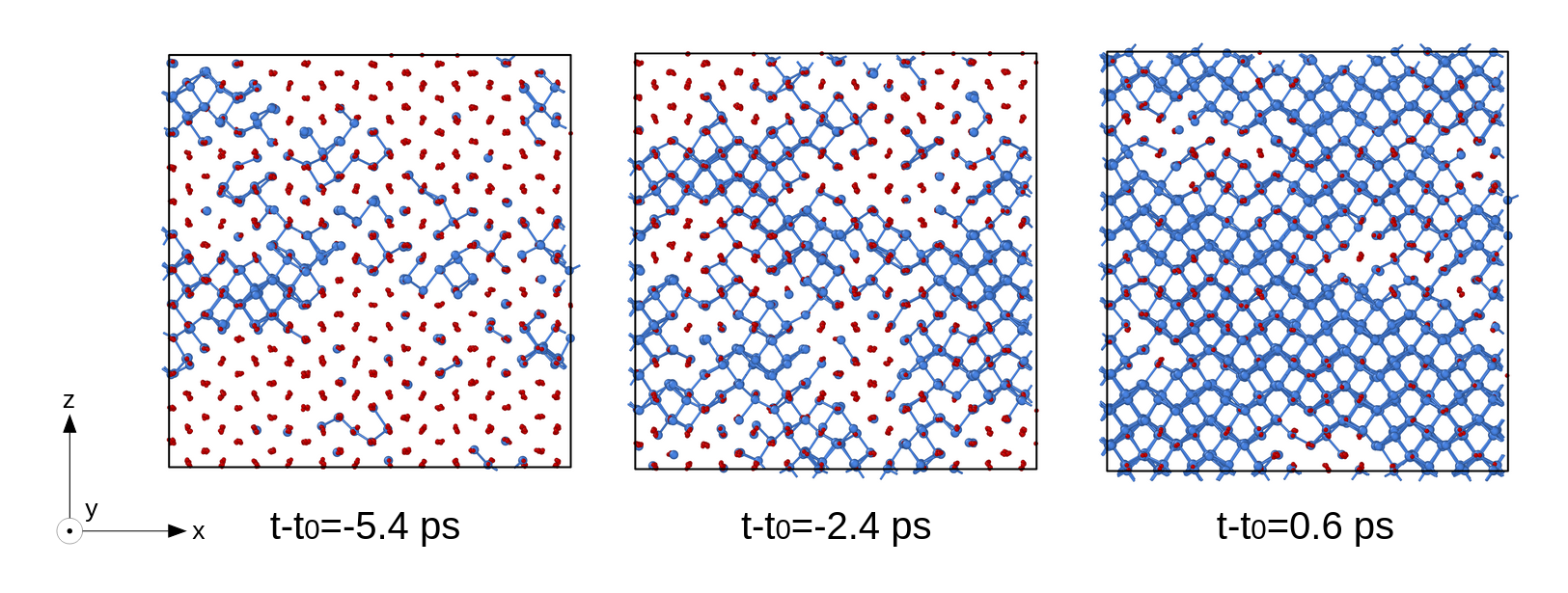}
    \caption{Snapshots of the phase transition from ZIF-4-cp-II to ZIF-4-cp. Only Zn ions are represented and they are classified using the SOAP-based classifier, color-coded depending on the predicted phase (ZIF-4-cp: blue, ZIF-4-cp-II: red).}
    \label{fig:CPII_to_CP_1st_simulation}
\end{figure}

\section{BPSF-based classifier results for CP-CPII interconversions}

\begin{figure}[H]
    \centering
    \includegraphics[width=0.6\linewidth]{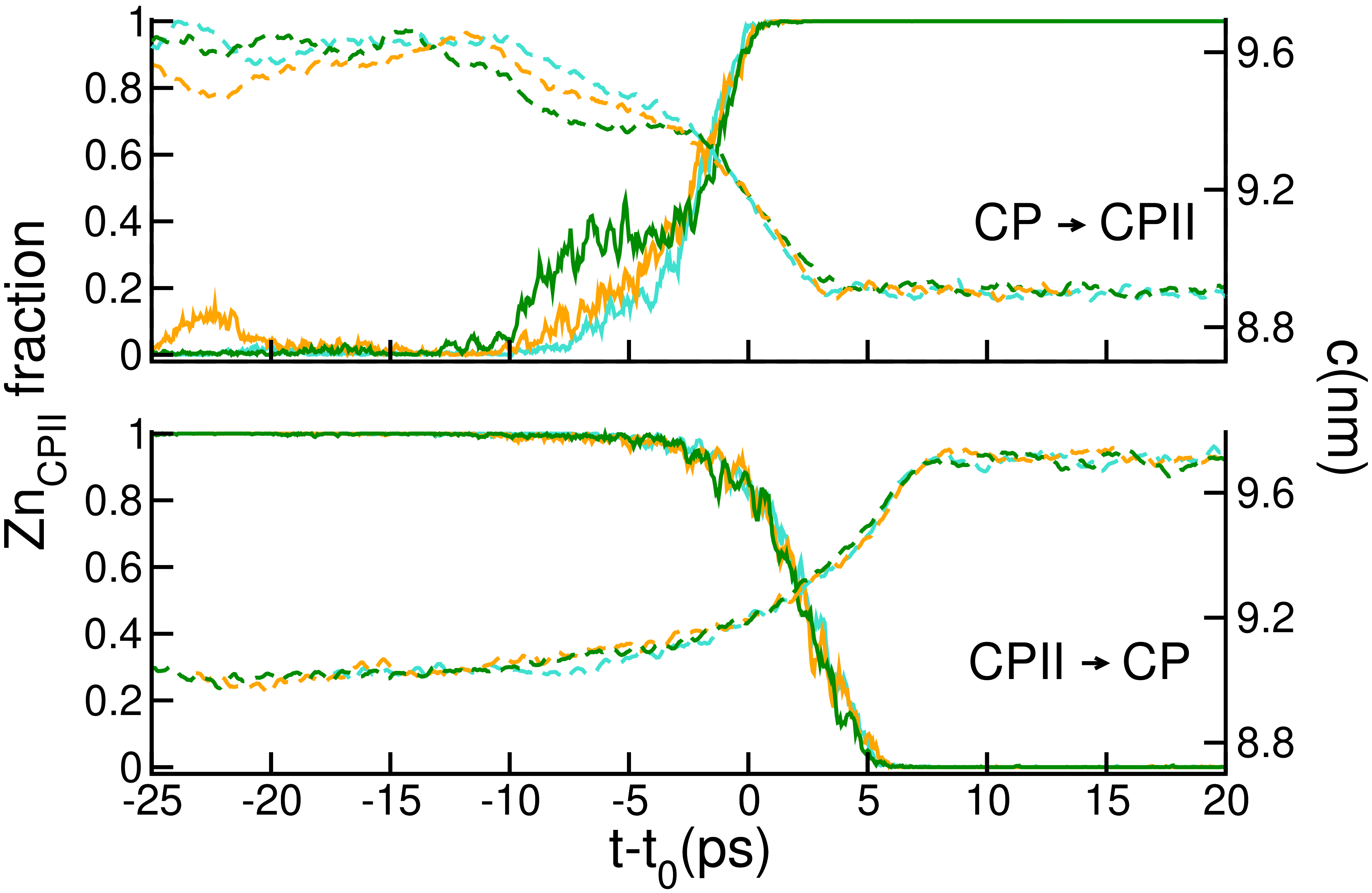}
    \caption{Fraction of Zn ions classified as CPII by the BPSF-based algorithm (left axis, full lines) and simulation box size in $z$ direction ($c$) (right axis, dashed lines) vs time for CP to CPII (upper panel) and CPII to CP (lower panel) transitions. All the trajectories are aligned at $t_0$, which corresponds to the time at which half of the Zn ions are classified as CPII. The different colors correspond to independent transition trajectories.}
    \label{fig:predictionsSI}
\end{figure}

\begin{figure}[H]
\begin{center}
\includegraphics[width=0.6\textwidth]{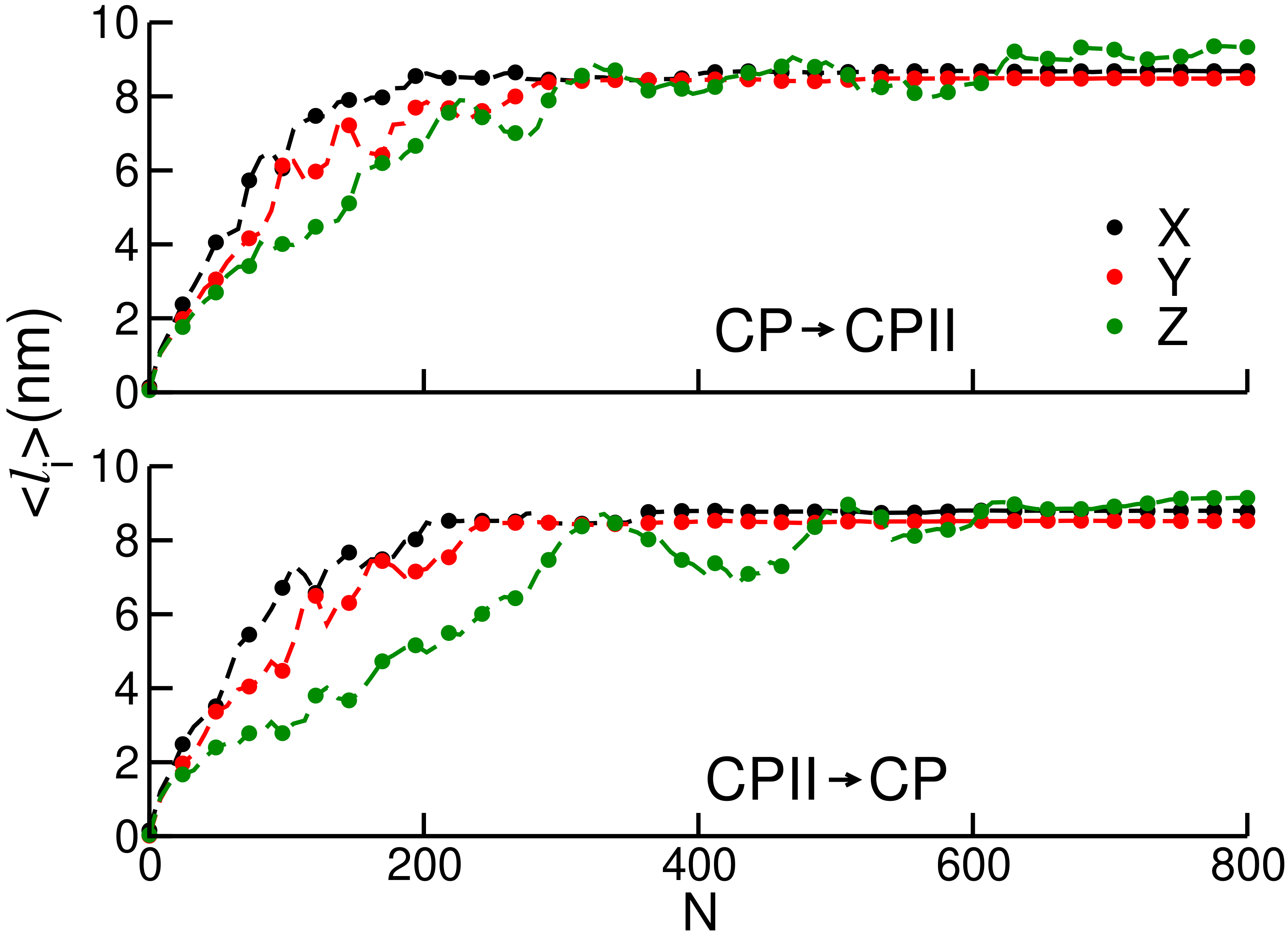}
\caption{Average cluster size $\langle l_i \rangle$ in $i$=X (black), Y (red) and Z (green) directions versus number of Zn ions comprising the cluster (N) for the BPSF-based classifier. Upper panel: CP to CPII transitions. Lower panel: CPII to CP transitions. The atoms that form the clusters correspond to the ones classified as belonging to the final phase in each case. Results from three independent simulations were averaged.} 
\label{fig:sizesSI}
\end{center}
\end{figure}

\begin{figure}[H]
\begin{center}
\includegraphics[width=0.6\textwidth]{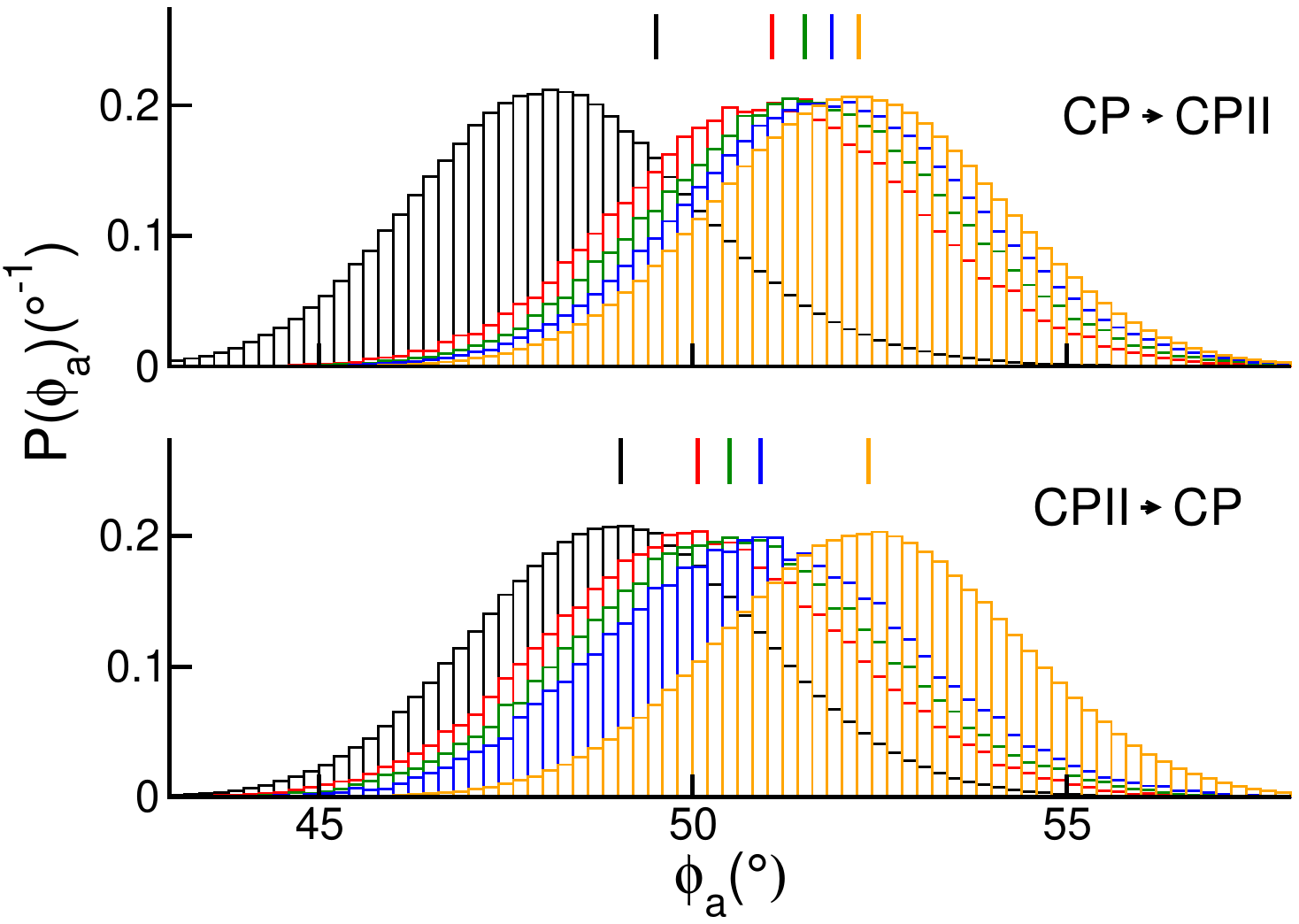}
\caption{Histograms of the angle $\phi_a$ between the vector normal to the surface formed by the four-membered rings of Zn atoms and the vector $\vec{a}$ that corresponds to the $x$ axis of the simulation box. A sketch of the angle $\phi_a$ is depicted in the upper right corner, where each ball corresponds to a Zn atom forming a 4-membered ring, the vertical stick indicates the normal vector to the surface and the arrowed stick depicts the vector $\vec{a}$. Zn atoms are colored according to the classification result of the BPSF-based classifier, blue for CP and red for CPII.
The histograms are divided by the number of Zn ions in the ring classified as CP. Mean values of each histogram are marked with vertical lines following the same color code.
Results obtained from transition trajectories of CP to CPII are plotted in the upper panel and the opposite ones in the lower panel.} 
\label{fig:tetragonsSI}
\end{center}
\end{figure}

\begin{figure}[H]
\begin{center}
\includegraphics[width=0.6\textwidth]{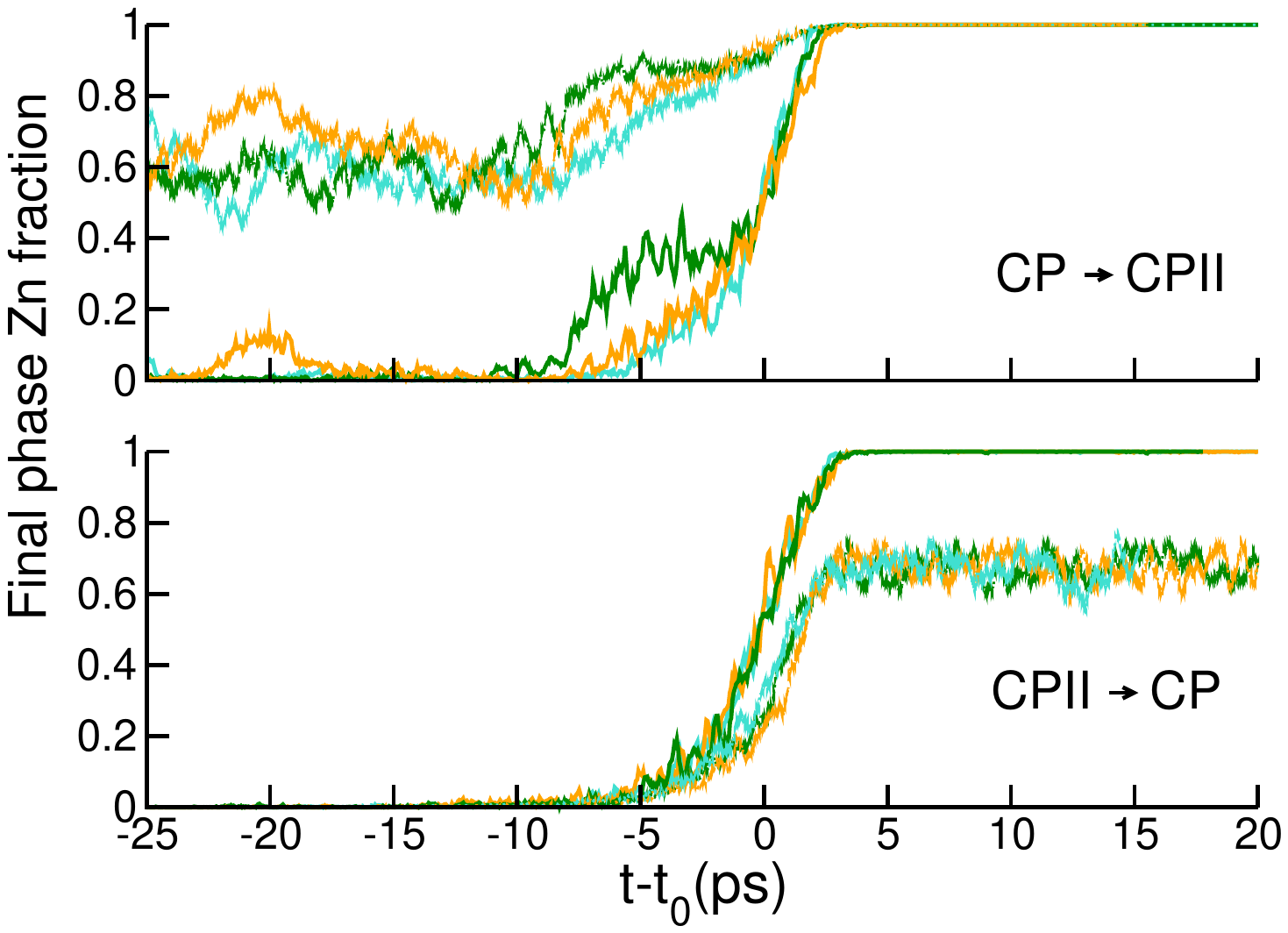}
\caption{BPSF classification output without CP data augmentation (dotted lines) and with data augmentation (full lines) for CP to CPII and CPII to CP transitions. }
\label{fig:predictions2}
\end{center}
\end{figure}

\begin{figure}[H]
\begin{center}
\includegraphics[width=0.6\textwidth]{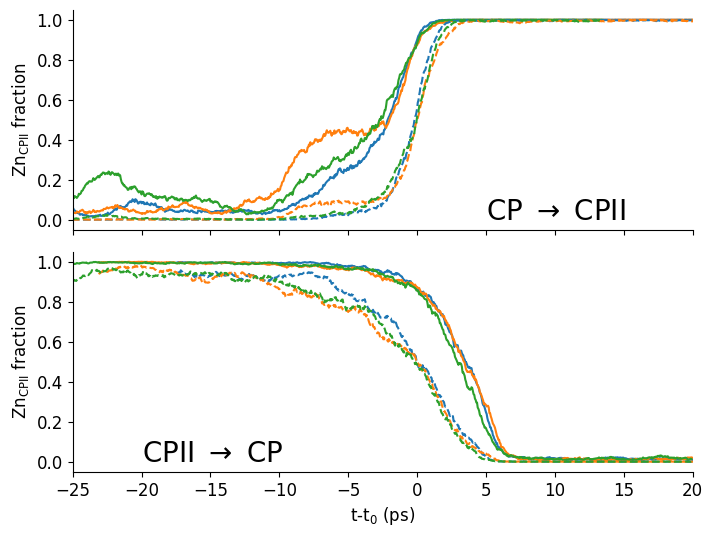}
\caption{SOAP classification output without CP data augmentation (full lines) and with data augmentation (dotted lines) for CP to CPII and CPII to CP transitions.  }
\label{fig:soap_without_cp_added}
\end{center}
\end{figure}

\section{Experimental phase diagram of ZIF-4}

\begin{figure}[H]
    \centering
    \includegraphics[width=0.65\linewidth]{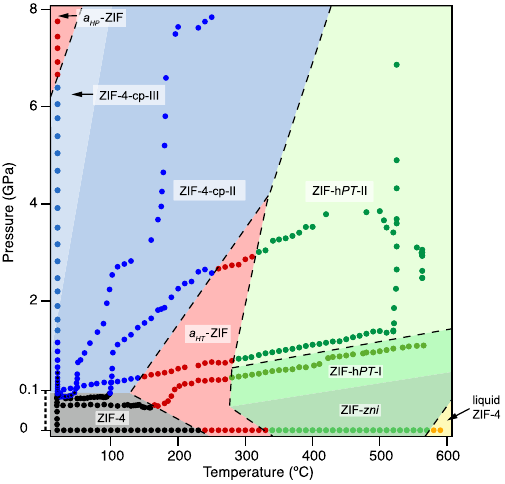}
    \caption{Experimental phase diagram of ZIF-4. Reproduced from Widmer et al.\cite{doi:10.1021/jacs.9b03234} under the Creative Commons Attribution (CC-BY) License.}
    \label{fig:phasediagram}
\end{figure}

\bibliography{article}

@article{10.1063/5.0128656,
    author = {Balestra, S. R. G. and Semino, R.},
    title = {Computer simulation of the early stages of self-assembly and thermal decomposition of ZIF-8},
    journal = {J. Chem. Phys.},
    volume = {157},
    number = {18},
    pages = {184502},
    year = {2022},
    issn = {0021-9606},
    doi = {10.1063/5.0128656},
    url = {https://doi.org/10.1063/5.0128656},
}

@Article{D3DD00236E,
author ="Castel, Nicolas and André, Dune and Edwards, Connor and Evans, Jack D. and Coudert, François-Xavier",
title  ="Machine learning interatomic potentials for amorphous zeolitic imidazolate frameworks",
journal  ="Digit. Discov.",
year  ="2024",
volume  ="3",
issue  ="2",
pages  ="355-368",
doi  ="10.1039/D3DD00236E",
url  ="http://dx.doi.org/10.1039/D3DD00236E",
}

@misc{batatia_mace_2023,
	title = {{MACE}: Higher Order Equivariant Message Passing Neural Networks for Fast and Accurate Force Fields},
	shorttitle = {{MACE}},
	url = {http://arxiv.org/abs/2206.07697},
	doi = {10.48550/arXiv.2206.07697},
	urldate = {2025-08-26},
	author = {Batatia, Ilyes and Kovács, Dávid Péter and Simm, Gregor N. C. and Ortner, Christoph and Csányi, Gábor},
	year = {2023},
	note = {arXiv:2206.07697},
}

@article{batatia2022mace,
  title={MACE: Higher order equivariant message passing neural networks for fast and accurate force fields},
  author={Batatia, Ilyes and Kovacs, David P and Simm, Gregor and Ortner, Christoph and Cs{\'a}nyi, G{\'a}bor},
  journal={Advances in neural information processing systems},
  volume={35},
  pages={11423--11436},
  year={2022}
}

@article{castel_challenges_2022,
	title = {Challenges in {Molecular} {Dynamics} of {Amorphous} {ZIFs} {Using} {Reactive} {Force} {Fields}},
	volume = {126},
	issn = {1932-7447},
	url = {https://doi.org/10.1021/acs.jpcc.2c06305},
	doi = {10.1021/acs.jpcc.2c06305},
	number = {45},
	urldate = {2025-08-26},
	journal = {J. Phys. Chem. C},
	author = {Castel, Nicolas and Coudert, François-Xavier},
	year = {2022},
	pages = {19532--19541},
}

@article{mendez_microscopic_2024,
	title = {Microscopic mechanism of thermal amorphization of {ZIF}-4 and melting of {ZIF}-zni revealed via molecular dynamics and machine learning techniques},
	volume = {12},
	issn = {2050-7496},
	url = {https://pubs.rsc.org/en/content/articlelanding/2024/ta/d3ta07361k},
	doi = {10.1039/D3TA07361K},
	language = {en},
	number = {8},
	urldate = {2025-08-26},
	journal = {J. Mater. Chem. A},
	author = {Méndez, Emilio and Semino, Rocio},
	year = {2024},
	pages = {4572--4582},
}

@article{D4TA05026F,
author ="Méndez, Emilio and Semino, Rocio",
title  ="Phase diagram of ZIF-4 from computer simulations",
journal  ="J. Mater. Chem. A",
year  ="2024",
volume  ="12",
issue  ="45",
pages  ="31108-31115",
doi  ="10.1039/D4TA05026F",
url  ="http://dx.doi.org/10.1039/D4TA05026F",
}

@article{doi:10.1021/jacs.9b03234,
author = {Widmer, Remo N. and Lampronti, Giulio I. and Chibani, Siwar and Wilson, Craig W. and Anzellini, Simone and Farsang, Stefan and Kleppe, Annette K. and Casati, Nicola P. M. and MacLeod, Simon G. and Redfern, Simon A. T. and Coudert, Fran{\c{c}}ois-Xavier and Bennett, Thomas D.},
title = {Rich Polymorphism of a Metal--Organic Framework in Pressure--Temperature Space},
journal = {J. Am. Chem. Soc.},
volume = {141},
number = {23},
pages = {9330-9337},
year = {2019},
doi = {10.1021/jacs.9b03234},
}

@misc{plimpton_lammps_2023,
	title = {{LAMMPS}: {Large}-scale {Atomic}/{Molecular} {Massively} {Parallel} {Simulator}},
	shorttitle = {{LAMMPS}},
	url = {https://zenodo.org/records/10806852},
	urldate = {2025-08-26},
	author = {Plimpton, Steven J. and Kohlmeyer, Axel and Thompson, Aidan P. and Moore, Stan G. and Berger, Richard},
	year = {2023},
	doi = {10.5281/zenodo.10806852},
}

@misc{LAMMPS_KOKKOS,
  doi = {10.48550/ARXIV.2508.13523},
  url = {https://arxiv.org/abs/2508.13523},
  author = {Johansson,  Anders and Weinberg,  Evan and Trott,  Christian R. and McCarthy,  Megan J. and Moore,  Stan G.},
  title = {LAMMPS-KOKKOS: Performance Portable Molecular Dynamics Across Exascale Architectures},
  year = {2025},
  copyright = {arXiv.org perpetual,  non-exclusive license}
}

@article{KOKKOS3,
  author={Trott, Christian R. and Lebrun-Grandié, Damien and Arndt, Daniel and Ciesko, Jan and Dang, Vinh and Ellingwood, Nathan and Gayatri, Rahulkumar and Harvey, Evan and Hollman, Daisy S. and Ibanez, Dan and Liber, Nevin and Madsen, Jonathan and Miles, Jeff and Poliakoff, David and Powell, Amy and Rajamanickam, Sivasankaran and Simberg, Mikael and Sunderland, Dan and Turcksin, Bruno and Wilke, Jeremiah},
  journal={IEEE Transactions on Parallel and Distributed Systems},
  title={Kokkos 3: Programming Model Extensions for the Exascale Era},
  year={2022},
  volume={33},
  number={4},
  pages={805-817},
  doi={10.1109/TPDS.2021.3097283}
}

@article{Bartok2013,
  title = {On representing chemical environments},
  volume = {87},
  ISSN = {1550-235X},
  url = {http://dx.doi.org/10.1103/PhysRevB.87.184115},
  DOI = {10.1103/physrevb.87.184115},
  number = {18},
  pages = {184115},
  journal = {Phys. Rev. B},
  publisher = {American Physical Society (APS)},
  author = {Bartók,  Albert P. and Kondor,  Risi and Csányi,  Gábor},
  year = {2013},
  month = may 
}

@article{scikit-learn,
  title={Scikit-learn: Machine Learning in {P}ython},
  author={Pedregosa, F. and Varoquaux, G. and Gramfort, A. and Michel, V.
          and Thirion, B. and Grisel, O. and Blondel, M. and Prettenhofer, P.
          and Weiss, R. and Dubourg, V. and Vanderplas, J. and Passos, A. and
          Cournapeau, D. and Brucher, M. and Perrot, M. and Duchesnay, E.},
  journal={J. Mach. Learn. Res.},
  volume={12},
  pages={2825--2830},
  year={2011}
}

@article{Triestram2026,
  title = {Identifying phase transitions in zeolitic imidazolate frameworks: microscopic insight from molecular simulations},
  volume = {17},
  ISSN = {2041-6539},
  url = {http://dx.doi.org/10.1039/D5SC09468B},
  DOI = {10.1039/d5sc09468b},
  number = {13},
  journal = {Chem. Sci.},
  publisher = {Royal Society of Chemistry (RSC)},
  author = {Triestram,  Léna and Coudert,  Fran\c{c}ois-Xavier},
  year = {2026},
  pages = {6734–6745}
}

@article{Aulakh2015,
  title = {The Importance of Polymorphism in Metal–Organic Framework Studies},
  volume = {54},
  ISSN = {1520-510X},
  url = {http://dx.doi.org/10.1021/acs.inorgchem.5b01311},
  DOI = {10.1021/acs.inorgchem.5b01311},
  number = {17},
  journal = {Inorg. Chem.},
  publisher = {American Chemical Society (ACS)},
  author = {Aulakh,  Darpandeep and Varghese,  Juby R. and Wriedt,  Mario},
  year = {2015},
  month = aug,
  pages = {8679–8684}
}

@article{Coudert2025,
  title = {Recent advances in stimuli-responsive framework materials: Understanding their response and searching for materials with targeted behavior},
  volume = {539},
  ISSN = {0010-8545},
  url = {http://dx.doi.org/10.1016/j.ccr.2025.216760},
  DOI = {10.1016/j.ccr.2025.216760},
  journal = {Coord. Chem. Rev.},
  publisher = {Elsevier BV},
  author = {Coudert,  Fran\c{c}ois-Xavier},
  year = {2025},
  month = sep,
  pages = {216760}
}

@article{Schneemann2014,
  title = {Flexible metal–organic frameworks},
  volume = {43},
  ISSN = {1460-4744},
  url = {http://dx.doi.org/10.1039/C4CS00101J},
  DOI = {10.1039/c4cs00101j},
  number = {16},
  journal = {Chem. Soc. Rev.},
  publisher = {Royal Society of Chemistry (RSC)},
  author = {Schneemann,  A. and Bon,  V. and Schwedler,  I. and Senkovska,  I. and Kaskel,  S. and Fischer,  R. A.},
  year = {2014},
  month = may,
  pages = {6062–6096}
}

@article{Peh2020,
  title = {Multiscale Design of Flexible Metal–Organic Frameworks},
  volume = {2},
  ISSN = {2589-5974},
  url = {http://dx.doi.org/10.1016/j.trechm.2019.10.007},
  DOI = {10.1016/j.trechm.2019.10.007},
  number = {3},
  journal = {Trends Chem.},
  publisher = {Elsevier BV},
  author = {Peh,  Shing Bo and Karmakar,  Avishek and Zhao,  Dan},
  year = {2020},
  month = mar,
  pages = {199–213}
}

@article{Hughes2012,
  title = {Thermochemistry of Zeolitic Imidazolate Frameworks of Varying Porosity},
  volume = {135},
  ISSN = {1520-5126},
  url = {http://dx.doi.org/10.1021/ja311237m},
  DOI = {10.1021/ja311237m},
  number = {2},
  journal = {J. Am. Chem. Soc.},
  publisher = {American Chemical Society (ACS)},
  author = {Hughes,  James T. and Bennett,  Thomas D. and Cheetham,  Anthony K. and Navrotsky,  Alexandra},
  year = {2012},
  month = dec,
  pages = {598–601}
}

@article{Bumstead2020,
  title = {Investigating the melting behaviour of polymorphic zeolitic imidazolate frameworks},
  volume = {22},
  ISSN = {1466-8033},
  url = {http://dx.doi.org/10.1039/D0CE00408A},
  DOI = {10.1039/d0ce00408a},
  number = {21},
  journal = {CrystEngComm},
  publisher = {Royal Society of Chemistry (RSC)},
  author = {Bumstead,  Alice M. and Ríos Gómez,  María Laura and Thorne,  Michael F. and Sapnik,  Adam F. and Longley,  Louis and Tuffnell,  Joshua M. and Keeble,  Dean S. and Keen,  David A. and Bennett,  Thomas D.},
  year = {2020},
  pages = {3627–3637}
}

@article{Smirnova2023,
  title = {Precise control over gas-transporting channels in zeolitic imidazolate framework glasses},
  volume = {23},
  ISSN = {1476-4660},
  url = {http://dx.doi.org/10.1038/s41563-023-01738-3},
  DOI = {10.1038/s41563-023-01738-3},
  number = {2},
  journal = {Nature Mater.},
  publisher = {Springer Science and Business Media LLC},
  author = {Smirnova,  Oksana and Hwang,  Seungtaik and Sajzew,  Roman and Ge,  Lingcong and Reupert,  Aaron and Nozari,  Vahid and Savani,  Samira and Chmelik,  Christian and Reithofer,  Michael R. and Wondraczek,  Lothar and K\"{a}rger,  J\"{o}rg and Knebel,  Alexander},
  year = {2023},
  month = dec,
  pages = {262–270}
}

@article{Hou2020,
  title = {Metal–organic framework gels and monoliths},
  volume = {11},
  ISSN = {2041-6539},
  url = {http://dx.doi.org/10.1039/C9SC04961D},
  DOI = {10.1039/c9sc04961d},
  number = {2},
  journal = {Chem. Sci.},
  publisher = {Royal Society of Chemistry (RSC)},
  author = {Hou,  Jingwei and Sapnik,  Adam F. and Bennett,  Thomas D.},
  year = {2020},
  pages = {310–323}
}

@article{Cheetham2018,
  title = {Thermodynamic and Kinetic Effects in the Crystallization of Metal–Organic Frameworks},
  volume = {51},
  ISSN = {1520-4898},
  url = {http://dx.doi.org/10.1021/acs.accounts.7b00497},
  DOI = {10.1021/acs.accounts.7b00497},
  number = {3},
  journal = {Acc. Chem. Res.},
  publisher = {American Chemical Society (ACS)},
  author = {Cheetham,  Anthony K. and Kieslich,  G. and Yeung,  H. H.-M.},
  year = {2018},
  month = feb,
  pages = {659–667}
}

@article{CastilloBlas2024,
  title = {Thermally activated structural phase transitions and processes in metal–organic frameworks},
  volume = {53},
  ISSN = {1460-4744},
  url = {http://dx.doi.org/10.1039/D3CS01105D},
  DOI = {10.1039/d3cs01105d},
  number = {7},
  journal = {Chem. Soc. Rev.},
  publisher = {Royal Society of Chemistry (RSC)},
  author = {Castillo-Blas,  Celia and Chester,  Ashleigh M. and Keen,  David A. and Bennett,  Thomas D.},
  year = {2024},
  pages = {3606–3629}
}

@article{Robertson2025,
  title = {Changes in the Long-Range Order and Local Atomic Structure of Zeolitic Imidazolate Frameworks under Extreme Conditions},
  volume = {65},
  ISSN = {1520-510X},
  url = {http://dx.doi.org/10.1021/acs.inorgchem.5c03812},
  DOI = {10.1021/acs.inorgchem.5c03812},
  number = {1},
  journal = {Inorg. Chem.},
  publisher = {American Chemical Society (ACS)},
  author = {Robertson,  Georgina P. and Anzellini,  Simone and Meneghini,  Carlo and Herlihy,  Anna and Amboage,  Monica and Keen,  David A. and Irifune,  Tetsuo and Bennett,  Thomas D.},
  year = {2025},
  month = dec,
  pages = {156–164}
}

@article{Vervoorts2021,
  title = {Structural Chemistry of Metal–Organic Frameworks under Hydrostatic Pressures},
  volume = {3},
  ISSN = {2639-4979},
  url = {http://dx.doi.org/10.1021/acsmaterialslett.1c00250},
  DOI = {10.1021/acsmaterialslett.1c00250},
  number = {12},
  journal = {ACS Mater. Lett.},
  publisher = {American Chemical Society (ACS)},
  author = {Vervoorts,  Pia and Stebani,  Julia and Méndez,  Alba S. J. and Kieslich,  Gregor},
  year = {2021},
  month = oct,
  pages = {1635–1651}
}

@article{Song2022,
  title = {Tuning the High‐Pressure Phase Behaviour of Highly Compressible Zeolitic Imidazolate Frameworks: From Discontinuous to Continuous Pore Closure by Linker Substitution},
  volume = {61},
  ISSN = {1521-3773},
  url = {http://dx.doi.org/10.1002/anie.202117565},
  DOI = {10.1002/anie.202117565},
  number = {21},
  pages = {e202117565},
  journal = {Angew. Chem. Int. Ed.},
  publisher = {Wiley},
  author = {Song,  Jianbo and Pallach,  Roman and Frentzel‐Beyme,  Louis and Kolodzeiski,  Pascal and Kieslich,  Gregor and Vervoorts,  Pia and Hobday,  Claire L. and Henke,  Sebastian},
  year = {2022},
  month = mar 
}

@article{Park2006,
  title = {Exceptional chemical and thermal stability of zeolitic imidazolate frameworks},
  volume = {103},
  ISSN = {1091-6490},
  url = {http://dx.doi.org/10.1073/pnas.0602439103},
  DOI = {10.1073/pnas.0602439103},
  number = {27},
  journal = {Proc. Nat. Acad. Sci.},
  publisher = {Proceedings of the National Academy of Sciences},
  author = {Park,  Kyo Sung and Ni,  Zheng and C\^oté,  Adrien P. and Choi,  Jae Yong and Huang,  Rudan and Uribe-Romo,  Fernando J. and Chae,  Hee K. and O’Keeffe,  Michael and Yaghi,  Omar M.},
  year = {2006},
  month = jul,
  pages = {10186–10191}
}

@article{BousselduBourg2014,
  title = {Thermal and mechanical stability of zeolitic imidazolate frameworks polymorphs},
  volume = {2},
  ISSN = {2166-532X},
  url = {http://dx.doi.org/10.1063/1.4904818},
  DOI = {10.1063/1.4904818},
  number = {12},
  pages = {124110},
  journal = {APL Mater.},
  publisher = {AIP Publishing},
  author = {Bouëssel du Bourg,  Lila and Ortiz,  Aurélie U. and Boutin,  Anne and Coudert,  Fran\c{c}ois-Xavier},
  year = {2014},
  month = dec 
}

@article{Helfrecht2019,
  title = {A new kind of atlas of zeolite building blocks},
  volume = {151},
  ISSN = {1089-7690},
  url = {http://dx.doi.org/10.1063/1.5119751},
  DOI = {10.1063/1.5119751},
  number = {15},
  pages = {154112},
  journal = {J. Chem. Phys.},
  publisher = {AIP Publishing},
  author = {Helfrecht,  Benjamin A. and Semino,  Rocio and Pireddu,  Giovanni and Auerbach,  Scott M. and Ceriotti,  Michele},
  year = {2019},
  month = oct 
}

@article{Behler2011,
  title = {Atom-centered symmetry functions for constructing high-dimensional neural network potentials},
  volume = {134},
  ISSN = {1089-7690},
  url = {http://dx.doi.org/10.1063/1.3553717},
  DOI = {10.1063/1.3553717},
  number = {7},
  pages = {074106},
  journal = {J. Chem. Phys.},
  publisher = {AIP Publishing},
  author = {Behler,  J\"{o}rg},
  year = {2011},
  month = feb 
}

@article{AndarziGargari2025,
  title = {Unveiling ZIF-8 Nucleation Mechanisms through Molecular Simulation: Role of Temperature,  Solvent,  and Reactant Concentration},
  volume = {37},
  ISSN = {1520-5002},
  url = {http://dx.doi.org/10.1021/acs.chemmater.5c02028},
  DOI = {10.1021/acs.chemmater.5c02028},
  number = {23},
  journal = {Chem. Mater.},
  publisher = {American Chemical Society (ACS)},
  author = {Andarzi Gargari,  Sahar and Semino,  Rocio},
  year = {2025},
  month = nov,
  pages = {9460–9470}
}

@article{Mendez2025,
  title = {Thermodynamic insights into the self-assembly of zeolitic imidazolate frameworks from computer simulations},
  volume = {16},
  ISSN = {2041-6539},
  url = {http://dx.doi.org/10.1039/D5SC00807G},
  DOI = {10.1039/d5sc00807g},
  number = {26},
  journal = {Chem. Sci.},
  publisher = {Royal Society of Chemistry (RSC)},
  author = {Méndez,  Emilio and Semino,  Rocio},
  year = {2025},
  pages = {11979–11988}
}

@article{Henke2018,
  title = {Pore closure in zeolitic imidazolate frameworks under mechanical pressure},
  volume = {9},
  ISSN = {2041-6539},
  url = {http://dx.doi.org/10.1039/C7SC04952H},
  DOI = {10.1039/c7sc04952h},
  number = {6},
  journal = {Chem. Sci.},
  publisher = {Royal Society of Chemistry (RSC)},
  author = {Henke,  Sebastian and Wharmby,  Michael T. and Kieslich,  Gregor and Hante,  Inke and Schneemann,  Andreas and Wu,  Yue and Daisenberger,  Dominik and Cheetham,  Anthony K.},
  year = {2018},
  pages = {1654–1660}
}

@article{larsen_robust_2016,
	title = {Robust structural identification via polyhedral template matching},
	volume = {24},
	issn = {0965-0393},
	url = {https://doi.org/10.1088/0965-0393/24/5/055007},
	doi = {10.1088/0965-0393/24/5/055007},
	language = {en},
	number = {5},
	urldate = {2026-04-03},
	journal = {Model. Simul. Mater. Sci. Eng.},
	publisher = {IOP Publishing},
	author = {Larsen, Peter Mahler and Schmidt, Søren and Schiøtz, Jakob},
	month = may,
	year = {2016},
	pages = {055007},
}

@article{rosset_signatures_2025,
	title = {Signatures of paracrystallinity in amorphous silicon from machine-learning-driven molecular dynamics},
	volume = {16},
	copyright = {2025 The Author(s)},
	issn = {2041-1723},
	url = {https://www.nature.com/articles/s41467-025-57406-4},
	doi = {10.1038/s41467-025-57406-4},
	language = {en},
	number = {1},
	urldate = {2026-04-03},
	journal = {Nature Commun.},
	publisher = {Nature Publishing Group},
	author = {Rosset, Louise A. M. and Drabold, David A. and Deringer, Volker L.},
	month = mar,
	year = {2025},
	pages = {2360},
}

@article{Chaplais2018,
  title = {Impacts of the Imidazolate Linker Substitution (CH\textsubscript{3},  Cl,  or Br) on the Structural and Adsorptive Properties of ZIF-8},
  volume = {122},
  ISSN = {1932-7455},
  url = {http://dx.doi.org/10.1021/acs.jpcc.8b08706},
  DOI = {10.1021/acs.jpcc.8b08706},
  number = {47},
  journal = {J. Phys. Chem. C},
  publisher = {American Chemical Society (ACS)},
  author = {Chaplais,  Gérald and Fraux,  Guillaume and Paillaud,  Jean-Louis and Marichal,  Claire and Nouali,  Habiba and Fuchs,  Alain H. and Coudert,  Fran\c{c}ois-Xavier and Patarin,  Joël},
  year = {2018},
  month = oct,
  pages = {26945–26955}
}

@article{Tanaka2014,
  title = {Adsorption-Induced Structural Transition of ZIF-8: A Combined Experimental and Simulation Study},
  volume = {118},
  ISSN = {1932-7455},
  url = {http://dx.doi.org/10.1021/jp500931g},
  DOI = {10.1021/jp500931g},
  number = {16},
  journal = {J. Phys. Chem. C},
  publisher = {American Chemical Society (ACS)},
  author = {Tanaka,  Hideki and Ohsaki,  Shuji and Hiraide,  Shotaro and Yamamoto,  Daigo and Watanabe,  Satoshi and Miyahara,  Minoru T.},
  year = {2014},
  month = apr,
  pages = {8445–8454}
}

@misc{kingma_adam_2017,
	title = {Adam: {A} {Method} for {Stochastic} {Optimization}},
	shorttitle = {Adam},
	url = {http://arxiv.org/abs/1412.6980},
	doi = {10.48550/arXiv.1412.6980},
	urldate = {2026-04-09},
	publisher = {arXiv},
	author = {Kingma, Diederik P. and Ba, Jimmy},
	month = jan,
	year = {2017},
	note = {arXiv:1412.6980 [cs]},
}

@misc{agarap_deep_2019,
	title = {Deep {Learning} using {Rectified} {Linear} {Units} ({ReLU})},
	url = {http://arxiv.org/abs/1803.08375},
	doi = {10.48550/arXiv.1803.08375},
	urldate = {2026-04-09},
	publisher = {arXiv},
	author = {Agarap, Abien Fred},
	month = feb,
	year = {2019},
	note = {arXiv:1803.08375 [cs]},
}

@article{jorgensen_interpretable_2026,
	title = {Interpretable visualizations of data spaces for classification problems},
	volume = {7},
	issn = {2632-2153},
	url = {https://doi.org/10.1088/2632-2153/ae466e},
	doi = {10.1088/2632-2153/ae466e},
	language = {en},
	number = {2},
	urldate = {2026-07-10},
	journal = {Machine Learning: Science and Technology},
	publisher = {IOP Publishing},
	author = {Jorgensen, Christian and Lin, Arthur Y and Vasavada, Rhushil and Cersonsky, Rose K},
	month = feb,
	year = {2026},
	pages = {025008},
}

@article{mcinnes_umap_2018,
	title = {{UMAP}: {Uniform} {Manifold} {Approximation} and {Projection}},
	volume = {3},
	issn = {2475-9066},
	shorttitle = {{UMAP}},
	url = {https://joss.theoj.org/papers/10.21105/joss.00861},
	doi = {10.21105/joss.00861},
	language = {en},
	number = {29},
	urldate = {2026-07-10},
	journal = {Journal of Open Source Software},
	author = {McInnes, Leland and Healy, John and Saul, Nathaniel and Großberger, Lukas},
	month = sep,
	year = {2018},
	pages = {861},
}

\end{document}